\documentclass[12pt, a4paper]{article}
\usepackage{graphicx, subcaption} 
\graphicspath{ {./images/} }
\usepackage{amsfonts, amsmath, amssymb, mathtools, amsthm, appendix, bbm}
\usepackage{natbib}
\usepackage{booktabs, float}
\usepackage{multirow, enumitem}
\usepackage{authblk}

\usepackage{geometry}
\usepackage{hyperref}
\hypersetup{
    colorlinks=true,
    linkcolor=blue,
    filecolor=blue,      
    urlcolor=blue,
    citecolor=blue
    }

\newcommand{\E}{\mathrm{E}}

\newcommand{\Y}{\mathbf{Y}}
\newcommand{\U}{\mathbf{U}}
\newcommand{\Z}{\mathbf{Z}}

\newcommand{\N}{\mathrm{N}}

\newcommand{\s}{\mathbf{\Sigma}}
\newcommand{\Om}{\mathbf{\Omega}}
\newcommand{\OUT}{\mathrm{spOUTAR}}

\newcommand{\D}{\mathbf{D}}
\newcommand{\A}{\mathbf{A}}
\newcommand{\IL}{\mathbf{L}}
\newcommand{\I}{\mathbf{I}}
\newcommand{\bY}{\mathbf{Y}}
\newcommand{\bU}{\mathbf{U}}

\newcommand{\bI}{\mathbf{I}}
\newcommand{\bA}{\mathbf{A}}
\newcommand{\bC}{\mathbf{C}}
\newcommand{\bd}{\mathbf{d}}

\newcommand{\bB}{\mathbf{B}}

\newcommand{\bL}{\mathbf{L}}
\newcommand{\bD}{\mathbf{D}}

\newcommand{\bphi}{\boldsymbol{\varphi}}
\newcommand{\bgm}{\boldsymbol{\gamma}}
\newcommand{\brho}{\boldsymbol{\rho}}

\newcommand{\bSigma}{\boldsymbol{\Sigma}}

\newcommand{\bOmega}{\boldsymbol{\Omega}}

\newcommand{\vect}{\mathrm{vec}}
\newcommand{\trans}{^{\mathrm{T}}}

\newtheorem{theorem}{Theorem}

\newcommand{\Frob}[1]{\Vert#1\Vert_{\mathrm{F}}}
\newcommand{\op}[1]{\Vert#1\Vert_{\mathrm{op}}}

\def\Diag{\hbox{Diag}}

\title{ Detecting Impact of the Great Recession on Macroeconomic Indicators}
\author[1]{Shuvrarghya Ghosh}
\author[2]{Arkaprava Roy}
\author[3]{Anindya Roy}
\author[1]{Subhashis Ghosal}
\affil[1]{Department of Statistics, North Carolina State University}
\affil[2]{Department of Biostatistics, University of Florida}
\affil[3]{Department of Mathematics and Statistics,\\ University of Maryland Baltimore County}

\begin{document}

\maketitle

\begin{abstract}
    \noindent We study structural changes in multivariate time series by estimating and comparing stationary graphs of macroeconomic time series before and after an economic crisis, such as the Great Recession. Building on a latent time series framework called  Orthogonally-rotated Univariate Time-series (OUT), we propose a shared-parameter framework-the spOUT autoregressive model (spOUTAR)-that jointly models two related multivariate time series and enables coherent Bayesian estimation of their corresponding stationary precision matrices. This framework provides a principled mechanism for detecting and quantifying which conditional relationships among the variables changed or formed following the crisis. Specifically, we study the impact of the Great Recession (December 2007–June 2009) that substantially disrupted global and national economies, prompting long-lasting shifts in macroeconomic indicators and their interrelationships. While many studies document its economic consequences, less is known about how the underlying conditional dependency structure among economic variables changed as economies moved from pre-crisis stability through the shock and back to normalcy. Using the proposed approach to analyze U.S. and OECD macroeconomic data, we demonstrate that spOUTAR effectively captures recession-induced changes in stationary graphical structure, offering a flexible and interpretable tool for studying structural shifts in economic systems.
\end{abstract}
\noindent%
{\bf Keywords:}  Conditional dependence, Langevin Monte Carlo, Macroeconomic Indicators, Stationary graph


\section{Introduction}
The Great Recession (December 2007--June 2009) had a profound impact on the global market in several ways. The Great Recession resulted in a prolonged employment slump that persisted long after its official end, as dated by the National Bureau of Economic Research. As unemployment surged, and housing prices and stock values collapsed, over 30 million people lost their jobs, with long-term unemployment doubling its previous record \citep{song2014long}. Household net worth fell significantly \citep{jacobsen2010us}. 
Consequently, numerous studies have examined economic outcomes, including the effects of unemployment on poverty, inequality, and earnings growth \citep{kalleberg2017us}.

In this paper, we focus on the impact of the Great Recession on the modification and moderation of the relationships between driving economic variables in the US economy \citep{kalleberg2017us} and, more broadly, in the economies of OECD (\textit{Organization for Economic Cooperation and Development}) countries \citep{ball2014long}.
We investigate changes in the structural associations among macroeconomic series in the U.S. economy, as well as cross-country associations in trade, the total economy, and several other key economic indicators. 

For stable economies and financial markets, representing implicit associations between variables in multiple economic/financial time series via stationary graphs can be useful and inferentially meaningful \citep{cordoni2024consistent}. However, when a systemic shock like the Great Recession hits an economy, markets and economic indicators can return to stability despite substantial changes to the series' stationary structure.  We study changes in the stationary graph among the associated variables before and after the Great Recession by estimating stationary precision matrices and comparing their forms. For our investigation, the economies are assumed to be stable before the recession and in the period following the event, once sufficient time has passed for them to return to normal.

We adopt the Orthogonally Rotated Univariate Time-series (OUT) model \citep{Roy2024}, a multivariate time-series framework, to estimate stationary graphs. However, we examine the expected change in the stationary graph in response to the economic shock. Thus, we are dealing with two potentially related multivariate time series and want to compare the stationary graphs associated with the two series; we propose a shared-parameter-OUT (spOUT) model. 
The univariate time series are modeled as autoregressive processes; thus, we refer to our final model as spOUTAR.
We apply this framework to estimate the stationary graphs before and after the crisis.
Consequently, our interest lies in learning the changes in the underlying stationary graphs. In this context, the variables denote vertices of an undirected graph, and an edge between two vertices denotes their conditional dependence given the other variables. Finally, the scientific goal is also to identify which conditional relationships have strengthened or weakened after the Great Recession.  We define these quantities within our Bayesian framework.

The remainder of the article is organized as follows. In Section \ref{sec:model}, we describe the spOUTAR model, which covers our choice of prior distributions and the specification of the posterior sampling strategy. Then, in Section \ref{sec:num}, we discuss the numerical results from simulation studies and real-data analyses. Finally, we conclude the paper in Section \ref{sec:diss}.

\section{Data and exploratory analysis}

\begin{figure}[h!]
    \centering
    \begin{subfigure}[b]{0.42\textwidth}
        \centering
        \includegraphics[width=\textwidth,trim=0.1cm 0.1cm 0.1cm 0.1cm,clip]{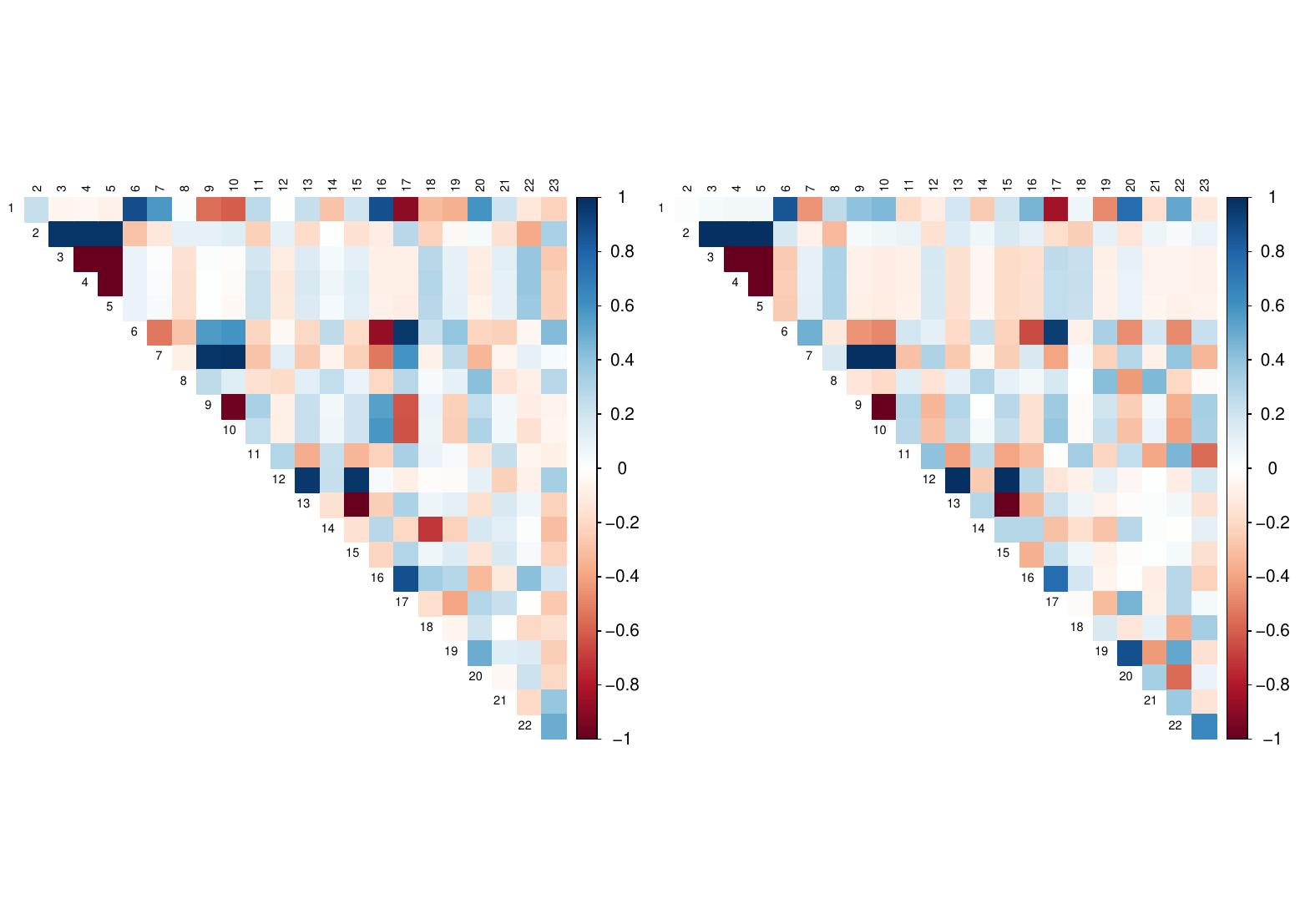}
        \caption{Partial correlations among NIPA variables: before (left) vs. after (right) the recession.}
        \label{fig:nipa_plot}
    \end{subfigure}
    \hfill
    \begin{subfigure}[b]{0.42\textwidth}
        \centering
        \includegraphics[width=\textwidth,trim=0.1cm 0.1cm 0.1cm 0.1cm,clip]{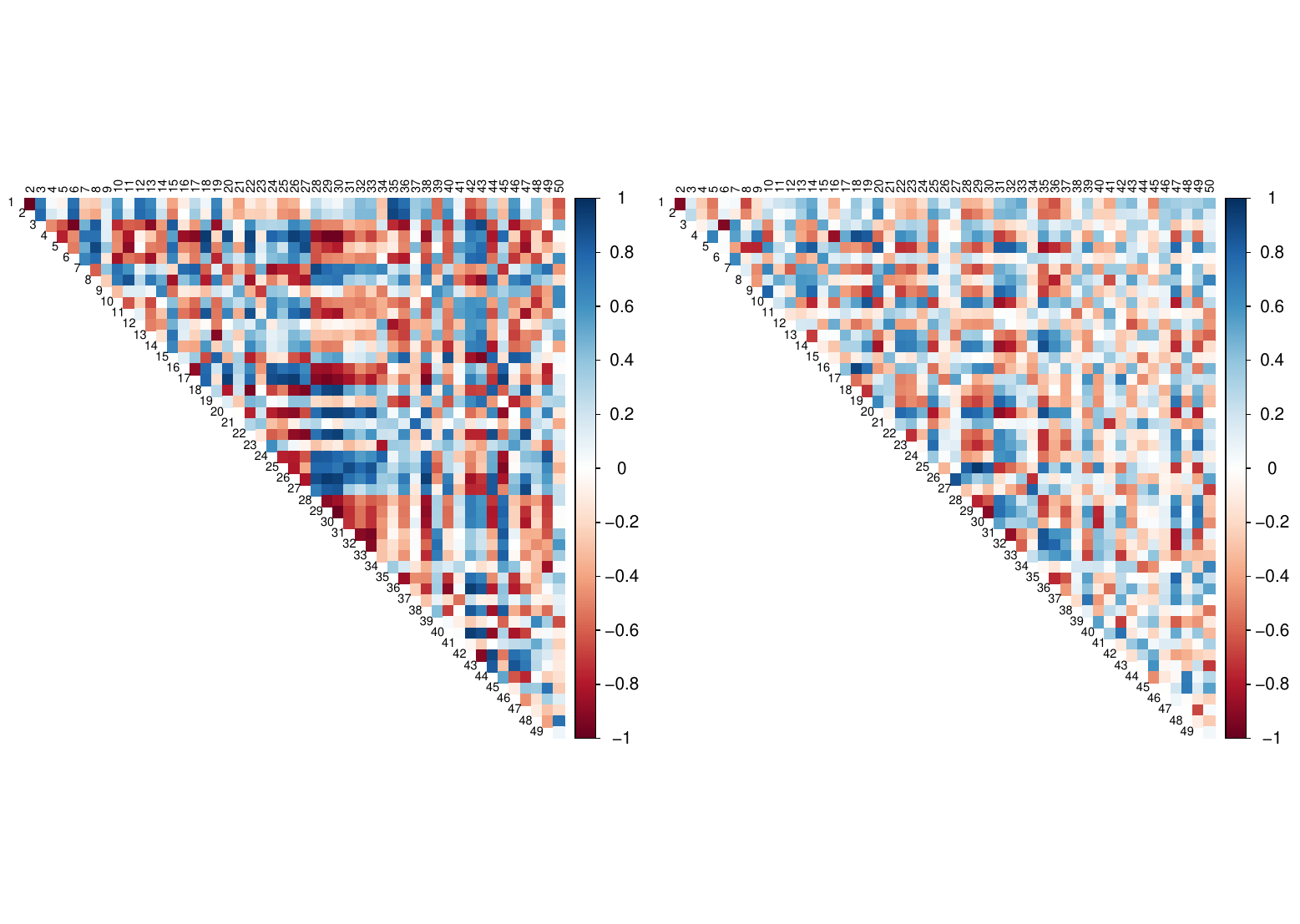}
        \caption{Partial correlations among employment and unemployment attributes: before (left) vs. after (right) the recession.}
        \label{fig:emp_plot}
    \end{subfigure}
    \caption{Changes in partial correlations between key macroeconomic indicators: (a) NIPA variables, and (b) employment/unemployment attributes, following the Great Recession.}
    \label{fig:nipa_emp_grid}
\end{figure}

In this article, we analyze two macroeconomic datasets. A brief description of each, along with some preliminary exploratory data analysis, is provided below.
\begin{figure}[h!]
    \centering
        \begin{subfigure}[b]{0.7\textwidth}
            \centering
            \includegraphics[width=\textwidth,trim=1.2cm 1cm 1cm 1cm,clip]{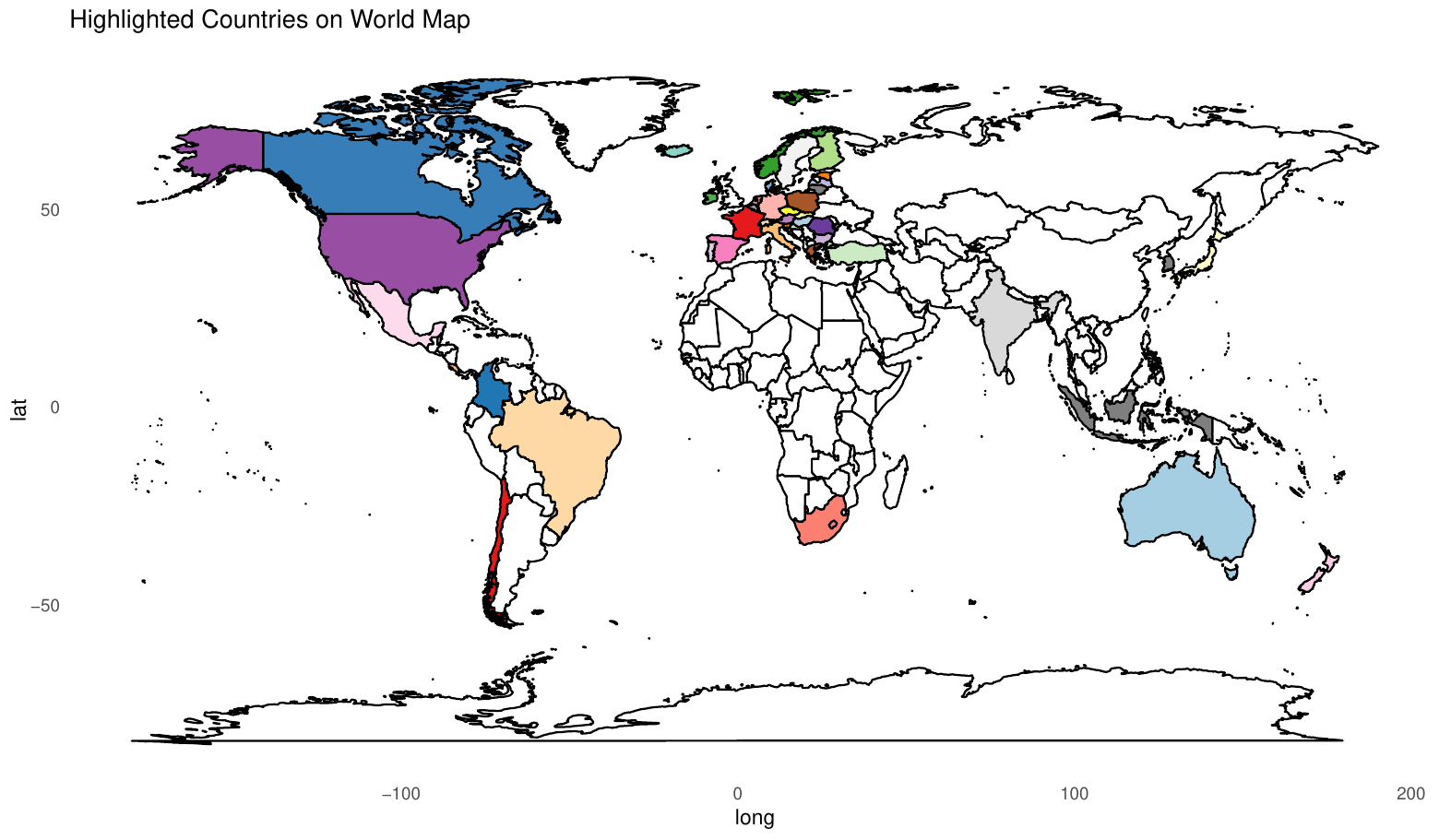}
        \end{subfigure}
    \caption{OECD countries. GDP and its components are studied for all the countries highlighted on the map.}
    \label{fig:oecd_world}
\end{figure}
\begin{enumerate}
    \item \textit{OECD nations GDP dataset}: The Organization for Economic Co-operation and Development (OECD) has 38 member countries, mainly consisting of developed nations with market-based economies. In this analysis, we use quarterly Gross Domestic Product (GDP) data from OECD member countries and seven additional nations (namely, Brazil, Bulgaria, Croatia, India, Indonesia, Romania and South Africa), totaling 45. We omit three non-OECD countries from our study, namely Russia, Saudi Arabia and Argentina, since the data for these countries are not available for the complete time period under investigation. The data, expressed in US dollars, is based on the expenditure approach, which computes GDP as the sum of the following components: final consumption expenditure of households and nonprofit institutions serving households (NPISH), final consumption expenditure of general government, gross fixed capital formation (i.e., investment), and net exports (exports minus imports). In addition to analyzing GDP, we examine each component individually. To develop an elementary understanding of pairwise dependencies among variables, while controlling for the effects of other variables and ignoring temporal dependence in each variable, we compute the partial correlation coefficients for every pair of variables in each data component. Two heatmaps illustrating the partial correlations among import trade variables and export trade variables are presented in Figure~\ref{fig:imp_plot} and Figure~\ref{fig:exp_plot}, respectively. The remaining plots are displayed in the Appendix.
    
    \begin{figure}[h!]
    \centering
    \begin{subfigure}[b]{0.49\textwidth}
        \centering
        \includegraphics[width=\textwidth,trim=0.1cm 0.1cm 0.1cm 0.1cm,clip]{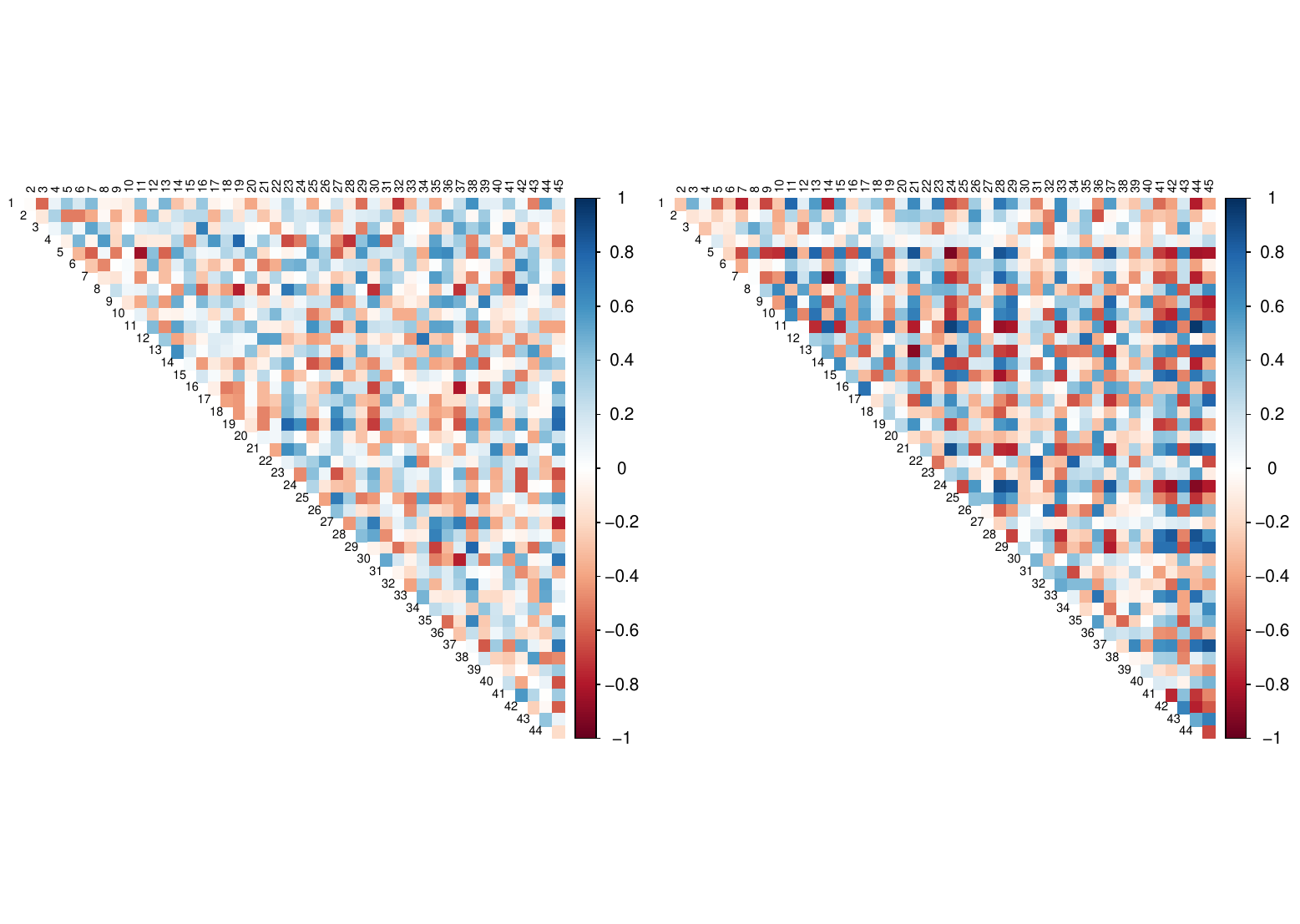}
        \caption{Partial correlations among import trade features: before (left) vs. after (right) the recession.}
        \label{fig:imp_plot}
    \end{subfigure}
    \hfill
    \begin{subfigure}[b]{0.49\textwidth}
        \centering
        \includegraphics[width=\textwidth,trim=0.1cm 0.1cm 0.1cm 0.1cm,clip]{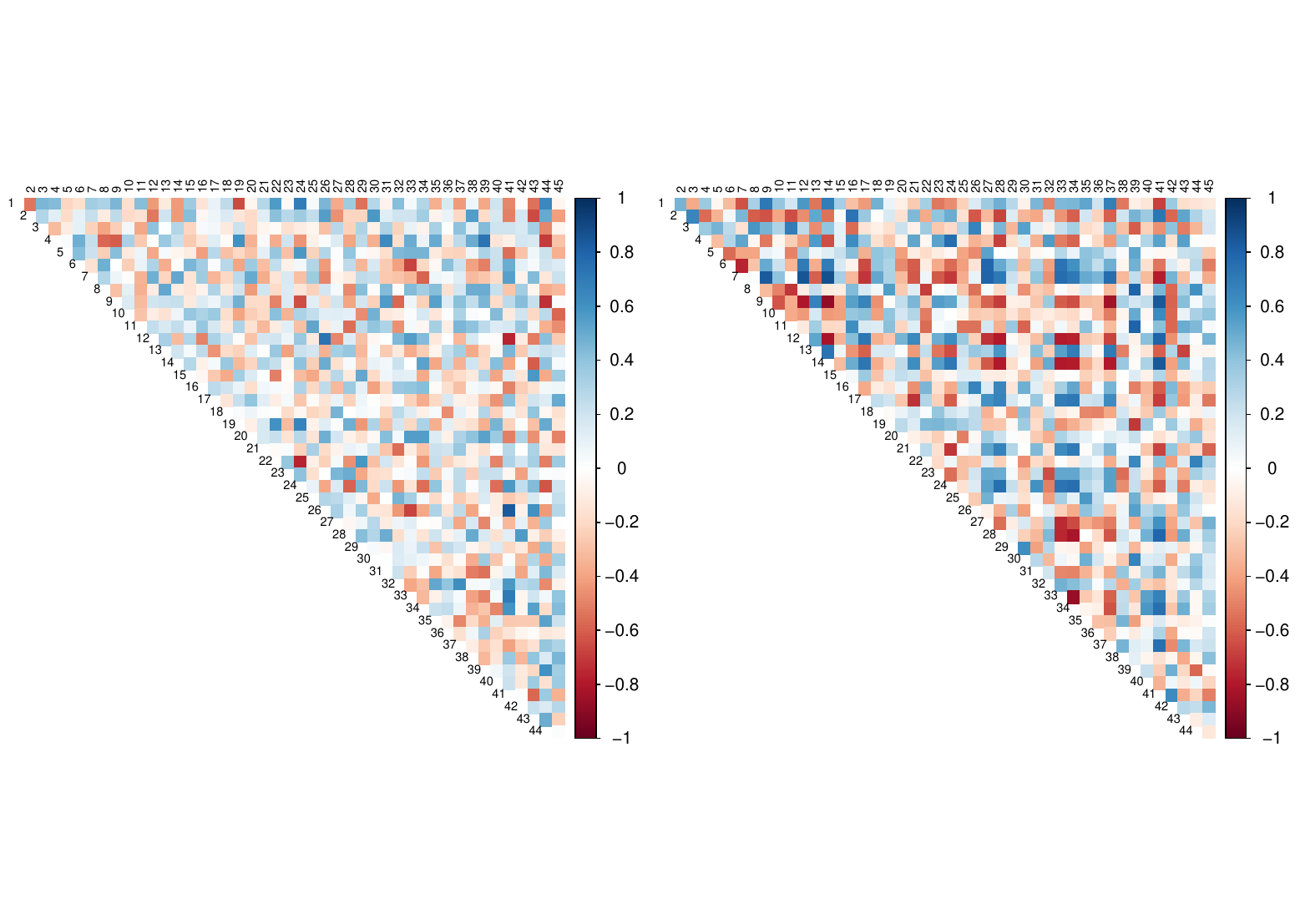}
        \caption{Partial correlations among export trade features: before (left) vs. after (right) the recession.}
        \label{fig:exp_plot}
    \end{subfigure}
    \caption{Changes in the partial correlations of (a) import trade features and (b) export trade features, following the Great Recession.}
    \label{fig:trade_corr_grid}
    \end{figure}

    \item \textit{FRED-QD macroeconomic dataset}: This quarterly macroeconomic data from the Federal Bank of St. Louis, comprising $248$ quarterly time series, closely resembles the Stock-Watson datasets (see \cite{stock2006, stock2012}), and is classified into $14$ groups. We consider $5$ groups for our analysis: National Income and Product Accounts (NIPA, $23$ variables); Industrial production ($16$ variables); Employment and unemployment ($50$ variables); Housing ($13$ variables); and Prices ($48$ variables). We preprocess the data as instructed in \cite{mccracken2020} and carry out the analysis in a manner similar to that outlined in Section~\ref{sec:OECD}. Consistent with the OECD data, we examine partial correlations for each component of this dataset. Heatmaps for the NIPA and Employment and Unemployment components are displayed in Figure \ref{fig:nipa_plot} and Figure \ref{fig:emp_plot}, respectively.

\end{enumerate}

The exploratory data analyses indicate appreciable changes in the partial correlation structure among our macroeconomic variables of interest, although they ignore the time-series nature of the data.
Since the two periods share the same set of macroeconomic variables, we propose a shared-parameter model tailored to study structural changes in paired multivariate time series.

\section{Methodology}\label{sec:model}

We write $[n]$ to denote the set $\{1, 2, \ldots, n\}$, for any $n \in \mathbb{N}$, the set of natural numbers. 
The notation $\vect:\mathbb{R}^{r \times s} \to \mathbb{R}^{rs}$ denotes the vectorization operator, a linear map that transforms a matrix $\mathbf{X} \in \mathbb{R}^{r\times s}$ to a ${rs \times 1}$-dimensional vector by stacking the columns of $\mathbf{X}$ on top of one another. Let $\|\mathbf{X}\|_{\mathrm{F}}$ denote the Frobenius norm of matrix $\mathbf{X}$ given by the relation $\|\mathbf{X}\|_{\mathrm{F}} = \sqrt{\sum_{i,j} X_{ij}^2}$. Let $\mathrm{diag}(x_1, \ldots, x_d)$ be a $d$-dimensional diagonal matrix with diagonal entries $\{x_1, \ldots, x_d\}$, and let $\boldsymbol{\sigma}=(\sigma_1,\ldots,\sigma_p)$ be the vector of innovation variances for the AR processes. Finally, we write $\N(\mu, \sigma^2)$ to denote a univariate normal distribution with mean $\mu \in \mathbb{R}$ and variance $\sigma^2$, $\sigma>0$; $\mathrm{Unif}(a, b)$ to denote a uniform distribution with finite support $(a, b) \subset \mathbb{R}$; $\mathrm{Logistic}(\mu, \sigma)$ to denote a logistic distribution with location $\mu \in \mathbb{R}$ and scale $\sigma>0$; $\mathrm{InvGamma}(\alpha, \beta)$ to denote an inverse gamma distribution with shape $\alpha > 0$ and scale $\beta > 0$; and $\mathrm{InvGauss}(\mu, \tau)$ to denote an inverse Gaussian distribution with mean $\mu > 0$ and shape $\tau > 0$. 

Consider a multivariate time series $\Y \in \mathbb{R}^{p \times (n_1+n_2)}$ with $p$ variables, each having $n_0$ observations, where it contains a period of disruption, such as the Great Depression and there are $n_1$ time points in period 1 (before the crisis) and $n_2$ time points in period 2 (after the crisis). Consequently, we form two data matrices $\Y_1 \in \mathbb{R}^{p \times n_1}$, and $\Y_2 \in \mathbb{R}^{p \times n_2}$. Let $\s_1$ and $\s_2$ denote the $p \times p$ dispersion matrices of $\Y_1$ and $\Y_2$ respectively. Then, using the framework in \cite{Roy2024}, $\Y_1$ and $\Y_2$ can be expressed as 
\[\Y_{1} = \s_1^{1/2}\U\Z_1 \text{  and  } \Y_{2} = \s_2^{1/2}\U\Z_2\] 
respectively. Since $\s_1$ and $\s_2$ are positive definite matrices, we write $\s_1^{1/2}$ and $\s_2^{1/2}$ to denote their respective Cholesky square roots. Let $\Z_1 \in\mathbb{R}^{p\times n_1}$ and $\Z_2 \in\mathbb{R}^{p\times n_2}$ be two matrices with $p$ independent latent variables, each representing a univariate time series with unit marginal variance. The variable $\U$ is a $p \times p$ orthogonal matrix that ensures order independence, as we use the Cholesky decomposition of the dispersion matrices to define the square root. 

We assume that the observations of the $j$th row of $\mathbf{Z}_1$ and $\mathbf{Z}_2$ are assumed to be part of the same univariate time series for all $j\in[p]$. However, because of the substantial gap between the two periods, they become independent yet retain the same underlying time-series characterization parameters. We assume that each univariate time series follows an autoregressive (AR) process for all $j\in[p]$. The orders of the AR processes may differ across univariate processes. For simplicity, unless otherwise stated, we assume a uniform order $q$ for all throughout this article. 

Let the $t_1$th time point of $\Y_1$ be expressed as $\Y_{1, t_1} = \s_1^{1/2}\U\Z_{1, t_1}$, for any $t_1 \in [n_1]$. Based on the relationship between $\Y_1$ and $\Z_1$, it is immediate that $\Z_{1, t_1} = \U^{\mathrm{T}}\Om_1^{1/2}\Y_{1, t_1}$, where $\Om_1 = \s_1^{-1}$. Since $\Om_1$ is a symmetric positive definite matrix, we can write $\Om_1 = (\I-\IL_1)\D_1^2(\I-\IL_1)^{\mathrm{T}}$, where $\D_1$ is a diagonal matrix and $\IL_1$ is a strictly lower-triangular matrix. Consequently, its Cholesky square root is $\Om_1^{1/2} = (\I-\IL_1)\D_1$. Similarly, for the $t_2$th time point of $\Y_2$, $\Z_{2, t_2} = \U^{\mathrm{T}}\Om_2^{1/2}\Y_{2, t_2}$, for any $t_2 \in [n_2]$, $\Om_2 = \s_2^{-1}$, and $\Om_2^{1/2} = (\I-\IL_2)\D_2$. In our proposed method, we assume $\Om_1$ and $\Om_2$ share a common diagonal matrix such that $\D_1 = \D_2 = \D$, and differ only in $\IL_1$ and $\IL_2$ that control the underlying connectivity patterns. Since $\Om_1$ gives the conditional dependency structure before the crisis, and $\Om_2$ gives the structure post-crisis, their difference, \(\Om_2 - \Om_1\), reveals changes in the dependency structure following the crisis.

Let the coefficient vector of the $j$th AR process be $\bphi_j = (\varphi_{j,1}, \ldots, \varphi_{j,q})^{\mathrm{T}}$, $j\in[p]$. Then, the coefficient matrix for all the AR processes combined is denoted by $\Psi = (\bphi_1, \ldots, \bphi_p)^{\mathrm{T}}$. Similarly, let the autocorrelation functions (ACF) and partial autocorrelation functions (PACF) of the $j$th time series be respectively defined by $\bgm_j = (\gamma_{j,1}, \ldots, \gamma_{j,q})^{\mathrm{T}}$ and $\brho_j = (\rho_{j,1}, \ldots, \rho_{j,q})^{\mathrm{T}}$, $j\in [p]$. 

We assume that the individual processes are subject to stationarity and causality constraints. In other words, we parameterize the (stationary) AR process in terms of its PACF parameters $\brho_j=\{\rho_{j,k}\}_{k=1}^q$, with $\rho_{j,k} \in (-1, 1)$, for any $j\in[p]$ and all $k\in [q]$. Subsequently, we obtain the AR process parameters, including the innovation variances, by applying the Durbin-Levinson recursion algorithm and the Yule–Walker equations as follows.
To get the AR coefficients $\bphi_j=(\varphi_{j,1}, \ldots, \varphi_{j,q})$ from $\brho_j$, we apply the the Durbin-Levinson algorithm. Then the implied $\bgm_j$ is obtained from $\bphi_j$ using the Yule–Walker equations. Finally, the innovation variance is obtained by the relation $\sigma^2_j = 1 - \sum_{k=1}^q \gamma_{j,k}\varphi_{j,k}$, ensuring that the marginal variances are 1. We denote the vector of $p$ innovation standard deviations by $\boldsymbol{\sigma} = (\sigma_1, \ldots, \sigma_p)$.

We recall that the correspondence between the PACF parameters and the AR coefficients is one-to-one and smooth \citep{ansley1986}. The stationarity and causality of the latent AR processes require that all roots of the equation $\zeta_i(\omega|\varphi_{j,k}:k\in [q])=0$, where 
$\zeta_j(\omega|\varphi_{j,k}:k\in [q])= 1+\sum_{k=1}^{q}\varphi_{j,k} e^{- i k  \omega}$, $j\in [p]$, 
lie outside the unit circle $\{\omega: |\omega|=1\}$ in the complex plane $\mathbb{C}$.
Let $\Upsilon$ denote the algebraic variety of the roots, the corresponding set for the $j$th latent AR process in terms of the PACF parameters $\Upsilon'$: 
\begin{align}
\label{root}
\Upsilon'=\{(\rho_{j,k}:k\in [q]): (\omega_1,\ldots,\omega_q)\in \mathbb{C}^q \mbox{ are the roots of }\zeta_j(\omega|\varphi_{j,k}:k\in [q])=0\}. 
\end{align}
It also follows that $\Upsilon'$ is a compact subset of $(0,1)^{q}$ in view of the homeomorphism between the PACF coefficients and the AR coefficients. Then, ensuring that the PACF parameters stay away from $\Upsilon'$ guarantees that the roots of $\zeta_j(\omega|\varphi_{j,k}:k\in [q])=0$, $j\in [p]$, stay away from the unit circle. 

Furthermore, for modeling purposes, we assume that the white noise in each AR process is normally distributed, although the true data-generating mechanism may not be normal. Then, the $t$th observation of the $j$th component of $\Z_1$ is modeled as
\[\mathrm{Z}_{1, j, t} = \sum_{k=1}^q \varphi_{j, k} \mathrm{Z}_{1, j, t-k} + \varepsilon_j, \quad\text{for all } j \in [p]; \: t \in \{q+1, \ldots, s\}, \]
where $\varepsilon_j \sim \N(0, \sigma^2_j)$ are independent and identically distributed (i.i.d.), and the model above can be viewed as a linear regression model. We ignore the first $q$ observations in each time series. The corresponding log-likelihood for the time-series $Z_{1,j}$, up to an additive term not containing the parameter, is
\[ \log \pi(\bphi_j|\mathrm{Z}_{1,j}) = -\frac{1}{2\sigma_j^2} \sum_{t=q+1}^s \big(\mathrm{Z}_{1, j, t} - \sum_{k=1}^q \varphi_{j, k} \mathrm{Z}_{1, j, t-k}\big)^2, \quad i=[p]. \] 
Consequently, the joint log-likelihood of all $p$ AR processes in $\Z_1$ is
\begin{equation}\label{eq:loglik_z1}
    \log\pi(\Psi|\Z_1) = -\sum_{j=1}^p\frac{1}{2\sigma_j^2} \sum_{t=q+1}^{n_1} \big(\mathrm{Z}_{1, j, t} - \sum_{k=1}^q \varphi_{j, k} \mathrm{Z}_{1, j, t-k}\big)^2.
\end{equation}
Similarly, for $\Z_2$,
\begin{equation}\label{eq:loglik_z2}
    \log\pi(\Psi|\Z_2) = -\sum_{j=1}^p\frac{1}{2\sigma_j^2} \sum_{t=q+1}^{n_2} \big(\mathrm{Z}_{2, j, t} - \sum_{k=1}^q \varphi_{j, k} \mathrm{Z}_{2, j, t-k}\big)^2.
\end{equation}
To draw posterior samples of $\IL_1$, we use the expression in \eqref{eq:loglik_z1}, thus using the data from the pre-crisis period only. Similarly, for $\IL_1$, we use the expression in \eqref{eq:loglik_z2}, involving the data from the post-crisis period. For all other parameters, we use the combined log-likelihood obtained by adding the expressions in the preceding two displays. Further computational details are described in the following subsections. 

Since $\Om_1 = (\I-\IL_1)\D^{2}(\I-\IL_1)^{\mathrm{T}}$ and $\Om_2 = (\I-\IL_2)\D^{2}(\I-\IL_2)^{\mathrm{T}}$, we induce priors on $\Om_1$ and $\Om_2$ by placing independent priors on $\D$, $\IL_1$, and $\IL_2$. We restrict $\D$ to have positive diagonal entries only, and $\IL_1$ and $\IL_2$ are strictly lower-triangular by definition. We assume $\Om_1$ and $\Om_2$ to be sparse matrices and induce them by making $\IL_1$ and $\IL_2$ sparse. Our complete set of parameters is $\{\D,\IL_1,\IL_2,\U,\Psi\}$.

We take a Bayesian route by assigning priors to all the model parameters, as discussed in the next subsection.
The posterior distribution is then approximated using a Markov chain Monte Carlo (MCMC) algorithm. Finally, we use the MCMC samples to characterize both the magnitude and direction of changes in connectivity.



\subsection{Prior specification}
The prior specifications for the Cholesky parameters and the orthogonal matrix in this article are similar to those used in \citet{Roy2024}. However, we include the full prior specification in this paper for completeness. We can set $\bOmega$ as $(\bI-\bL) \bD^2 (\bI-\bL)\trans + T^{-a}\bI_p$, where $a>1/2$ is fixed, independent priors are specified for $\bD=\Diag(\bd)$ and $\bL$. Then writing $\bC=(\bI-\bL)\bD$, an approximate $\bOmega^{1/2}$ can be set as $\bB=\bC + \frac{1}{2}T^{-a}(\bC^{-1})^{\mathrm{T}}$. Specifically,
$\bB\bB^{T}=\bOmega+\frac{1}{4}T^{-2a}(\bC^{-1})^{\mathrm{T}}\bC^{-1}$. 
\begin{itemize}
\item \textit{Prior for $\D$}: Define $\D = \mathrm{diag}\left(d_{1}, \ldots, d_{p}\right)$ where each diagonal element is independently distributed as $\mathrm{InvGauss}(\xi, 1)$. This prior has an exponential-like tail near both zero and infinity. Lastly, we put a prior distribution $\N(0, \sigma_d^2)$ on $\xi$ and set $\sigma_d$ to some large constant to make the prior weakly-informative. 
  
  \item \textit{Prior for $\IL_1$ and $\IL_2$}: Each sub-diagonal component of $\IL_1$ is independently distributed as $H_{\lambda_{\IL_1}}(X)$ given $\lambda_{\IL}$, where $X \sim \mathrm{N}\left(0, \sigma_{\IL_1}^{2}\right), H_{\lambda_{\IL_1}}(x)=x \mathbbm{1}\{|x|>\lambda_{\IL_1}\}$ is the hard-thresholding operator at level $\lambda_{\IL_1}$ and $\lambda_{\IL_1} \sim \mathrm{Unif}\left(0, 100\right)$. Further, we put an $\mathrm{InvGamma}(0.01, 0.01)$ prior on $\sigma_{\IL_1}^2$. Similarly, we put priors on $\IL_2$, $\lambda_{\IL_2}$, and $\sigma_{\IL_2}^2$.
  
  \item \textit{Prior for $\U$}: Using the Cayley representation for the special orthogonal matrices, we can reparameterize  $\U=\left(\I-\A\right)\left(\I+\A\right)^{-1}$ where $\A$ is a skew-symmetric matrix. A skew-symmetric matrix $\A$ has all diagonal entries zero, and sub-diagonal entries are unrestricted. We assume $\A$ to be sparse and the sparsity of $\A$ is achieved by specifying a thresholding-based prior for the sub-diagonal entries similar to that used for $\IL_1$ and $\IL_2$. The sub-diagonal entries of $\A$ are independently distributed as $H_{\lambda_{\A}}(X)$ given $\lambda_{\A}$, where $X \sim \mathrm{N}\left(0, \sigma_{\A}^{2}\right), H_{\lambda_{\A}}(x)=x \mathbbm{1}\{|x|>\lambda_{\A}\}$ is the hard-thresholding operator at level $\lambda_{\A}$ and $\lambda_{\A} \sim \mathrm{Unif}\left(0, 100\right)$. We put an $\mathrm{InvGamma}(0.1, 0.1)$ prior on $\sigma_{\A}^2$. 

  The Cayley transformation maps skew-symmetric matrices onto a subset of \textit{special orthogonal group of dimension $p$}, denoted $SO(p)$, consisting of orthogonal matrices that do not have $(-1)$ as an eigenvalue. However, the resulting image set is dense in $SO(p)$, and thus the chosen prior is convenient enough for practical purposes.
  
  \item \textit{Prior for $\Psi$}: To maintain stationarity and causality constraints on the AR parameters, we put $\mathrm{Unif}(-1, 1)$ priors on the corresponding PACFs $\rho_{j,1}, \ldots, \rho_{j,q}$ for $j \in [q]$, and apply the Durbin-Levinson algorithm to map them back to the AR parameters and innovation variance assuming unit marginal variance. 
\end{itemize}

\subsection{Sampling from the posterior}

\textbf{Model initialization}: To start the MCMC chain from a reasonable starting value, we use a hot start in the following manner.

\begin{enumerate}
  \item To initialize $\Om_1$ and $\Om_2$, we use the precision matrix estimates $\hat{\Om}_1$ and $\hat{\Om}_2$ obtained from the output of the Gaussian graphical model (GGM; see \cite{Mohammadi2015}) fit to the pre-crisis data and post-crisis data, respectively.  The GGM model is implemented using the \texttt{bdgraph} package in \texttt{R} \citep{Mohammadi2019}. This step ignores the across-time dependence. Subsequently, we compute the modified Cholesky factorization of $\hat{\Om}_1$ to get $\hat{\IL}_1$ and $\hat{\D}_1$, and of $\hat{\Om}_2$ to get $\hat{\IL}_2$ and $\hat{\D}_2$.
  \item Initialize $\IL_1 = \hat{\IL}_1$, $\IL_2 = \hat{\IL}_2$, and $\D = (\hat{\D}_1 + \hat{\D}_2)/2$. Set $\A = \boldsymbol{0}$ so that $\U = \I$. Obtain $\Z_1$ using the relation $\Z_1 = \U^{\mathrm{T}}\D(\I-\IL_1)\Y_1$, and $\Z_2$ from the relation $\Z_2 = \U^{\mathrm{T}}\D(\I-\IL_2)\Y_2$.
  \item Initialize the AR coefficients of all $p$ AR processes using the \texttt{ar}() function in \texttt{R}. Compute the ACF and PACFs of the corresponding coefficients using \texttt{inla.ar.phi2acf}() and \texttt{inla.ar.phi2PACF}() functions from the \texttt{R} package \texttt{INLA} by \citet{Gomez2020}. Note that given the values for other initial parameters, we can obtain $\Z_t$ from $\Y_t$, and hence estimate the AR coefficients.
  \item Set $\lambda_\A = \lambda_{\IL_1} = \lambda_{\IL_2} = 0$. Initialize $\sigma^2_\A$, $\sigma^2_{\IL_1}$, and $\sigma^2_{\IL_2}$ using draws from the inverse gamma posterior based on initial values of $\mathrm{vec}(\A)$,  $\mathrm{vec}(\IL_1)$, and $\mathrm{vec}(\IL_2)$.
\end{enumerate}
\noindent
\textbf{MCMC sampling}: Considering the structure of the model and the prior specification, the posterior is not completely analytically tractable. Hence, we use Metropolis-Hastings (M-H) and Metropolis-adjusted Langevin Monte Carlo (MALA) algorithms to sample from the posterior. The AR coefficients are subject to causality and stationarity constraints, and hence we sample only the AR parameters for the first $1500$ iterations, keeping $\IL_1$, $\IL_2$, $\D$, and $\A$ fixed at their initial values. After the AR parameters have stabilized, we start sampling the values of $\IL_1$, $\IL_2$, $\D$, and $\A$ with $\lambda_\A$ and $\lambda_{\IL_1}, \lambda_{\IL_2}$ fixed to zero until another $1000$ iterations. Further, it is important to note that hard spike-and-slab priors on the entries of $\IL_1$, $\IL_2$, and $\A$ are computationally expensive. Therefore, during sampling, we use a computationally convenient modification of the prior via hard or soft thresholding, as described below.

For the matrix $\boldsymbol{A}$, we let a sub-diagonal entry $a_{i j}, i>j$, be distributed as $S_{\lambda^{\prime}}(X):=X(1-$ $\left.\lambda^{\prime} /|X|\right)_{+}$, where $X \sim \mathrm{N}\left(0, \sigma_{T}^{2}\right)$ and $\lambda^{\prime}$ is a hyperparameter, independently of other sub-diagonal entries. Then it follows that for all $i>j$,

$$
a_{ij} \lvert\, \lambda^{\prime} \sim 
\begin{cases}
\delta_{0}, &\text {with probability } \Phi\left(\frac{\lambda^{\prime}}{\sigma_{T}}\right)-\Phi\left(-\frac{\lambda^{\prime}}{\sigma_{T}}\right); \\ 
\phi\left(\frac{\left|a_{i j}\right|+\lambda^{\prime}}{\sigma_T} \operatorname{sign}\left(a_{i j}\right)\right), &\text{otherwise.}
\end{cases}
$$

Finally, we put a prior $\lambda^{\prime} \sim \mathrm{Unif}(\lambda_{l}, \lambda_{u})$. We set $\lambda_{l}=0$ and $\lambda_{u}$ to some large constant.\\
We use Gibbs sampling to sample $\D, \IL_1, \IL_2, \A$ and $\Psi$. To initialize the chain, we use a hot start based on the estimated parameter values.

\begin{itemize}[parsep=0pt, partopsep=0pt]
  \item \textit{Sampling the matrix} $\A$: We consider Metropolis-Hastings (M-H) moves for the lower-triangular entries $\bA$. Specifically, we propose a small additive movement in $\bA$ from its initial position, using realizations from a multivariate normal distribution with mean equal to the previously accepted value and covariance matrix equal to a scalar multiple of the identity. 
  
  \item \textit{Sampling the matrices $\IL_1$ and $\IL_2$}: We use Metropolis-adjusted Langevin algorithm (MALA) to update $\IL_1$ and $\IL_2$. MALA combines aspects of the M-H algorithm with Langevin dynamics, enhancing efficiency by leveraging gradient information of the target distribution. The algorithm updates sample positions by adding a drift term, based on the gradient of the logarithm of the target density, to a random Gaussian noise term. This gradient-guided proposal step encourages movement toward regions of higher probability, thereby reducing the random-walk behavior common in other MCMC methods. Each proposed sample is accepted or rejected based on the M-H criterion \citep{roberts2002}.
  
   \setlength{\parindent}{1.5em} Computing the gradient of the hard-thresholding operator is difficult due to its discontinuity. Thus, following \cite{Cai2020}, we approximate the thresholding operator by 
   $$\mathbbm{1}\{|x|>\lambda\} \approx\{1+2 \pi^{-1} \tan ^{-1}((x^{2}-\lambda^{2}) / h_{0}\} / 2,$$ 
   with $h_{0}$ chosen as a small number such as $10^{-8}$. This helps with gradient-based updates to $\IL_1$ and $\IL_2$. The gradient of the likelihood with respect to $\IL_1$ and $\IL_2$ can be found in Appendix \ref{app:l}.

  \item \textit{Sampling the matrix} $\D$: Likewise, as in the last step, we use a MALA routine to sample elements of $\D$. Due to the positivity constraint on diagonal elements of $\D$, we update this parameter in the log scale with the necessary Jacobian adjustment. The gradient of the likelihood with respect to $\D$ used to compute the Langevin diffusion is given in Appendix \ref{app:d}.
  
  \item \textit{Sampling the AR parameters}:  Note that the AR processes are independent and are parametrized in terms of PACFs, which are bounded in $(-1,1)$.
  
  \setlength{\parindent}{1.5em} We update these parameters using the M-H algorithm. Here, we again reparameterize the PACF parameters for $j$th AR process, $\brho_j$, as $\log((\brho_j + 1) / 2) = \textbf{V}_j$ (say). Then, the uniform prior on the components of $\bphi_j$ induces $\mathrm{Logistic}(0, 1)$ prior on the components of $\textbf{V}_j$ with unrestricted support. This enables the implementation of the adaptive M-H algorithm to update the $\textbf{V}_j$'s, as described in \cite{haario2001adaptive}. 
  To map the PACF parameter back to the AR parameters, the R function \texttt{inla.ar.PACF2phi}() from {\tt INLA} package will be used.
  
  \item \textit{Sampling the thresholding parameters}: We update $\lambda_{\IL_1}$ and $\lambda_{\IL_2}$ using a random walk M-H step in the log scale with a Jacobian adjustment every 20th iteration. The parameter $\lambda_\A$ is also updated similarly for every $30$th iteration. Update $\sigma_{\IL_1}^2$, $\sigma_{\IL_2}^2$ and $\sigma_\A^2$ accordingly using updated values of $\mathrm{vec}(\IL_1)$, $\mathrm{vec}(\IL_2)$ and $\mathrm{vec}(\A)$ using draws from the inverse gamma posterior distribution.

\end{itemize}

We draw a total of $10000$ samples and discard the first $5000$ as burn-in. In the first 1500 iterations, we update only the AR process parameters, keeping all others fixed at their initial values. Subsequently, we start updating $\IL_1$, $\IL_2$, $\D$, and $\A$, keeping the thresholding parameters set to zero for the next 1000 iterations. Then, from the $2500$th iteration onward, we update all parameters. Note that these adjustments are done during the burn-in phase of the MCMC. The above-mentioned iteration numbers are carefully selected through our extensive numerical experiments. Even at \(a=3\), which yields the largest nugget considered, \(T^{-a}\) is at most \(50^{-3}=8\times10^{-6}\) over the reported sample sizes. Numerical comparisons confirmed that including or omitting the nugget produced identical results to the reported precision. We therefore omit the nugget from the simulations for numerical simplicity.

For the M-H algorithm, the acceptance rate is maintained between 15--40\% by rescaling the magnitude of the variances of the proposals accordingly for every $100$th iteration to ensure adequate mixing. Similarly, for MALA, it is kept between 45--70\% by adjusting the step size parameter at every $100$th iteration. 

\noindent \textbf{Prediction:} Suppose that we want to predict the value of $\Y_T$, for some $T > n$. Following the burn-in period of the sampler, in the $i$th MCMC iteration of the AR process of order $q$, we sequentially compute 
\[\Y_{t,i}^{*} = \s_i^{1/2}\U_i\Z_{[t-q,t],i}\Psi_i + \boldsymbol{\sigma}_i\odot \boldsymbol{\varepsilon}_i , \quad t\in \{n+1, \ldots, T\},\] 
where $\boldsymbol{\varepsilon}_i$ are independent innovations with all components independent standard normal, and $\odot$ represents the Hadamard (componentwise) product of two vectors or matrices of the same size.  
Consequently, the posterior mean estimate of $\Y_T$ is given by $\hat{\Y}_T = \sum_i \Y_{T, i}^{*} / N$, where $N$ is the number of iterations of MCMC excluding the burn-in period. Clearly, for one-step-ahead predictions, we take $T=n+1$.

\noindent \textbf{Posterior inference on connectivity changes:} We first calculate the posterior samples of the differences in the precision matrices $\Om_{21}=\Om_2 - \Om_1$ (say) by computing this difference for each posterior sample and then determine the $95\%$ credible interval for each entry. An edge is classified as having a positive (negative) shift if the two nodes it connects share a stronger (weaker) relationship after the crisis. In practice, we identify an edge to have a positive (negative) shift if its credible interval lies entirely in the positive (negative) region of the real line. We will use blue lines to indicate positive shifts and red lines to indicate negative shifts in the graphs displayed in this section, the next section, and in the appendix. 
The adjacency matrix is then constructed by assigning edges to those node pairs for which the credible interval of the corresponding entry of $\Om_{21}$ excludes zero. 

\section{Posterior convergence}
\label{sec:convergence}

Now, we derive the posterior contraction rate $\epsilon_T$ for the parameter $\bOmega$ at its true value $\bOmega_0$, given the observation $\bY^{(T)}=\{\bY_t\}_{t=1}^{\mathrm{T}}$, namely, a sequence $\epsilon_T$ such that 
$$\E \Pi (\| \bOmega-\bOmega_0\|_2 >m_T \epsilon_T|\bY^{(T)})\to 0 \mbox{ for all }m_T\to \infty, \mbox{ 
as }T\to\infty.$$ 
We shall show that the rate for the i.i.d. case obtained by \cite{banerjee2015bayesian} prevails under mild conditions on the autoregressive parameters. The primary difference between their derivation and the present situation is that we must also incorporate the latent autoregressive process parameters and the orthogonal transform as nuisance parameters. Moreover, since we use the Whittle likelihood to simplify the likelihood, the model is misspecified, and hence, unlike the i.i.d. case of \cite{banerjee2015bayesian}, the general theory of posterior contraction rate in \cite{ghosal2017fundamentals} cannot be directly applied. Instead, we follow the approach of \cite{Roy2024}, who used a crucial contiguity property of the Whittle likelihood observed in \cite{choudhuri2004contiguity} and 
extended the general posterior contraction rate theorem for such a likelihood. The primary difference between the setup of \cite{Roy2024} and ours is that in the present situation, the latent processes are guided by parametric models of autoregressive processes instead of only having smooth spectral densities approximated by B-spline expansions. As a result, in their setting, the smoothness of the spectral densities affects the final rate, restricting the speed of convergence. Our situation is essentially a case of infinitely smooth spectral densities, and hence, the rate in the i.i.d. case is replicated. Nevertheless, we allow the orders of the latent autoregressive processes to be arbitrary, which necessitates more careful analysis than a simple parametric model. 

As in \cite{Roy2024}, the contraction rate can only be obtained in a weaker sense (see Theorem~3 of \cite{Roy2024}) without additional boundedness assumptions on the domains of the parameters. To obtain a contraction rate in terms of the parameter of interest $\bOmega$, additional conditions ensuring the boundedness of eigenvalues are needed, which necessitate truncating the priors for certain parameters used in modeling from their original form. This is needed for the technical derivation, which has little practical impact because the truncation domain is purely conceptual and may be arbitrarily large. Let $\lesssim$ stand for an inequality up to a constant multiple. 
We assume that the following conditions hold: 

\begin{description}
\item [(A1)] The true precision matrix $\bOmega_0$ has eigenvalues that are bounded and bounded away from zero and has $s$ non-zero sub-diagonal entries. 
\item [(A2)] Lower-dimensionality of rotation: The true orthogonal transformation matrix $\bU_0$ has Cayley representation $(\bI_p-\bA_0)(\bI_p+\bA_0)^{-1}$, where $\bA_0$ has at most $s$ non-zero sub-diagonal entries, all lying in a fixed bounded interval. 
\item [(A3)] Compact domains: 
\begin{itemize}
\item The diagonal $d_1,\ldots,d_p$ of the Cholesky decomposition lie within a fixed compact subinterval of $(0,\infty)$, and the prior densities are truncated to the corresponding domains; Also, $\op{\bI-\bL}\le C$ for some fixed constant $C$.
\item the collection of PACF parameters $(\rho_{j,k}: k\in [q])$ of the latent AR processes lies in a fixed compact subset of $(-1,1)^{q}\setminus \Upsilon'$, where $\Upsilon'$ is \eqref{root} and the uniform prior on the PACF parameters is truncated to such a subset. 
\end{itemize}
\item [(A4)] 
Growth of dimension: $\log p\lesssim \log T$. 
\item [(A5)] Prior positivity: The prior densities of the autoregressive parameters of the latent processes, as well as for $d_1,\ldots,d_p$, are continuous and positive at their true values. 
\end{description}

Let the subscript $0$ attached to a symbol stand for its true value. 
The following theorem characterizes the posterior contraction rate. 

\begin{theorem}
\label{thm:main}
Under Conditions {\rm (A1)--(A5)}, for every sequence $m_T\to \infty$, 
 $$\E \Pi(\Frob{\bSigma-\bSigma_0} \ge m_T \epsilon_T|\bY^{(T)})\to 0, \quad  
      \E \Pi(\Frob{\bOmega-\bOmega_0}\ge m_T \epsilon_T|\bY^{(T)})\to 0,$$
where $\epsilon_T=\sqrt{(p+s)(\log p)/T}$ and $s$ stands for the number of nonzero subdiagonal entries of $\bOmega_0$ and $\bA_0$.  
\end{theorem}

The proof of Theorem~\ref{thm:main}  consists of verifying the conditions of Theorem~\ref{thm:pseudoposterior general} of the Appendix and is deferred to the Appendix.

\section{Numerical studies}
\label{sec:num}
In this section, we present the results of the real-data analyses, evaluate the model's predictive performance, and assess the underlying graphical structure of the data across various simulation settings.

\subsection{Real data analysis}\label{sec:real}
We analyze the dynamic characteristics of the macroeconomic components of the Gross Domestic Product (GDP) series before and after the Great Recession. The Great Recession began in December 2007 and lasted until June 2009. To compare the behavior of the series before and after the Great Recession, we split each series into two parts, one before December 2007 and the other after July 2009. We use two real datasets that are updated in real time. In both cases, we also compare the one-step ahead predictive performance of our method with that of the sparse vector autoregressive model (VAR) (see \cite{lutkepohl2013, basu2015}), where the sparse VAR forecasts are obtained using the \texttt{sparsevar} R package. In addition, we illustrate the underlying graphical structure using the estimated precision matrices and discuss the resilience of connections between economic factors based on differences in the estimated marginal precisions before and after recession periods. Throughout the following sections, the sparse VAR model will be referred to as ``Sparse-VAR''.

\subsubsection{Quarterly GDP data of OECD nations}\label{sec:OECD}

We consider data from the quarter $1$ (Q$1$) of $1998$ to Q$4$ of $2019$, excluding data from the recession period $2007$ Q$4$ to $2009$ Q$3$. In each case, we model the first difference of the natural logarithm of the series to remove any long-term trend. Small scaling adjustments have been made to avoid numerical instabilities. Subsequently, we fit Sparse-VAR and spOUTAR models of orders $2$ through $8$, and present the root mean squared errors in Table~\ref{tab:oecd}.
For inference, we consider the estimated graphs based on the order that gives the best predictive performance for any given category.

\begin{table}[h!]
\centering
\caption{One-step prediction errors of Sparse-VAR and spOUTAR models for different orders of autoregressive processes for OECD-QD data.}
\resizebox{0.70\textwidth}{!}{%
\begin{tabular}{@{}lllllllll@{}}
\toprule
\multirow{2}{*}{GDP Component (USD)}                                                      &\multirow{2}{*}{Model}       & \multicolumn{7}{c}{q}                                                                                                   \\ \cmidrule(l){3-9} 
                                                                                          &    & 2                 & 3                 & 4                 & 5        & 6                 & 7                 & 8        \\ \midrule
\multirow{2}{*}{Total Economy}                                                            & Sparse-VAR   & 0.1210          & 0.1154          & 0.1233            & 0.1219 & 0.1275          & 0.1275          & 0.1275 \\ 
                                                                                          & spOUTAR & \textbf{0.0092}          & 0.0105          & 0.0098          & 0.0096 & 0.0096 & 0.0102          & 0.0099 \\ \midrule
\multirow{2}{*}{Import expenditure}                                                       & Sparse-VAR   & 1.5221          & 1.5372          & 1.5632          & 1.5656 & 1.4395          & 1.3874          & 1.0003 \\ 
                                                                                          & spOUTAR & 0.0738 & 0.0729          & 0.0714          & 0.0757 & 0.0696          & \textbf{0.0692}          & 0.0742 \\ \midrule
\multirow{2}{*}{Export expenditure}                                                       & Sparse-VAR   & 1.3968          & 1.3968          & 1.0550          & 1.3946 & 1.3968          & 1.3968          & 1.3968 \\ 
                                                                                          & spOUTAR & 0.0277          & \textbf{0.0262}          & 0.0272 & 0.0275 & 0.0271          & 0.0289          & 0.0269 \\ \midrule
\multirow{2}{*}{\begin{tabular}[c]{@{}l@{}}Total capital \\ expenditure\end{tabular}}     & Sparse-VAR   & 3.5345          & 3.5695          & 4.3061          & 2.8892 & 3.6046          & 2.8249          & 2.1092 \\ 
                                                                                          & spOUTAR & 0.1477          & \textbf{0.1467} & 0.1544          & 0.1633 & 0.1576          & 0.1653          & 0.1751 \\ \midrule
\multirow{2}{*}{\begin{tabular}[c]{@{}l@{}}General government\\ expenditure\end{tabular}} & Sparse-VAR   & 1.0361          & 0.6447          & 0.6457          & 0.6504  & 0.5276          & 0.4800          & 0.3851 \\ 
                                                                                          & spOUTAR & 0.0255          & 0.0222          & \textbf{0.0194} & 0.0229 & 0.0231          & 0.0211           & 0.0176  \\ \midrule
\multirow{2}{*}{\begin{tabular}[c]{@{}l@{}}Household \\ expenditure\end{tabular}}         & Sparse-VAR   & 0.1628            & 0.1599          & 0.1628            & 0.1628   & 0.1628            & 0.1521          & 0.1500 \\ 
                                                                                          & spOUTAR & 0.0160          & \textbf{0.0151}          & 0.0158          & 0.0158 & 0.0157          & 0.0155 & 0.0208 \\ \bottomrule
\end{tabular}%
}
\label{tab:oecd}
\end{table}

The prediction error for the 2019 Q4 time point is computed using only $\Om_2$. The spOUTAR model outperforms the Sparse-VAR model in all cases, indicating superior predictive performance. To construct the graph adjacency matrix for each case, we follow the approach discussed at the end of the previous subsection based on the posterior samples of ${\Om}_{21}=\Om_2-\Om_1$. Next, we compute the posterior mean of the difference, $\bar{\Om}_{21}$; extract the values corresponding to the edges in the adjacency matrix; sort these values in descending order; and report the top 10 edges in Table~\ref{tab:oecdcon}.

\begin{table}[h!]
\centering
\caption{Top 10 edges in terms of absolute value of conditional dependencies in OECD-QD data groups based on the sorted absolute differences in the posterior mean of $\Om_{21}$.}
\resizebox{\textwidth}{!}{%
\begin{tabular}{@{}lr@{}}
\toprule
Group                                                                    & Dependencies (egdes)                                                                              \\ \midrule
Total economy                                                            & (IDN, JPN), (SWE, JPN), (LVA, JPN), (LUX, IND), (AUS, GBR),
(AUS, CHE), (CAN, LTU), (ISR, JPN), (AUS, JPN), (AUS, CHL)     \\
Import expenditure                                                       & (COL, LTU), (SVK, NLD), (MEX, ITA), (BGR, NLD), (ITA, LVA),
(MEX, HUN), (BGR, IDN), (BGR, GBR), (LUX, LTU), (MEX, SVK)  \\
Export expenditure                                                       & (CRI, AUT), (CRI, HUN), (BGR, LUX), (ESP, HRV), (NZL, FIN),
(ESP, SWE), (FIN, COL), (IND, HRV), (BEL, ISR), (ESP, POL)    \\
\begin{tabular}[c]{@{}l@{}}Total capital \\ expenditure\end{tabular}     & (ITA, COL), (GRC, FIN), (FIN, IND), (NOR, PRT), (LVA, DEU),
(ISL, ESP), (IND, POL), (ESP, FIN), (LVA, CHE), (ITA, IND)  \\
\begin{tabular}[c]{@{}l@{}}General government\\ expenditure\end{tabular} & (BGR, CHE), (CHE, SVK), (SVK, HRV), (NZL, HRV), (NOR, IND),
(BEL, IND), (IND, HRV), (USA, BEL), (HRV, LTU), (BGR, KOR) \\
\begin{tabular}[c]{@{}l@{}}Household\\ expenditure\end{tabular} & (AUS, KOR), (USA, GBR), (GBR, SWE), (USA, AUS), (SWE, NLD),
(PRT, JPN), (BEL, SWE), (CZE, KOR), (IDN, JPN), (BGR, KOR)  \\ \bottomrule
\end{tabular}%
}
\label{tab:oecdcon}
\end{table}

\begin{figure}[h!]
    \centering
    \begin{subfigure}[b]{0.42\textwidth}
        \centering
        \includegraphics[width=\textwidth,trim=5.5cm 4.0cm 4.0cm 3.5cm,clip]{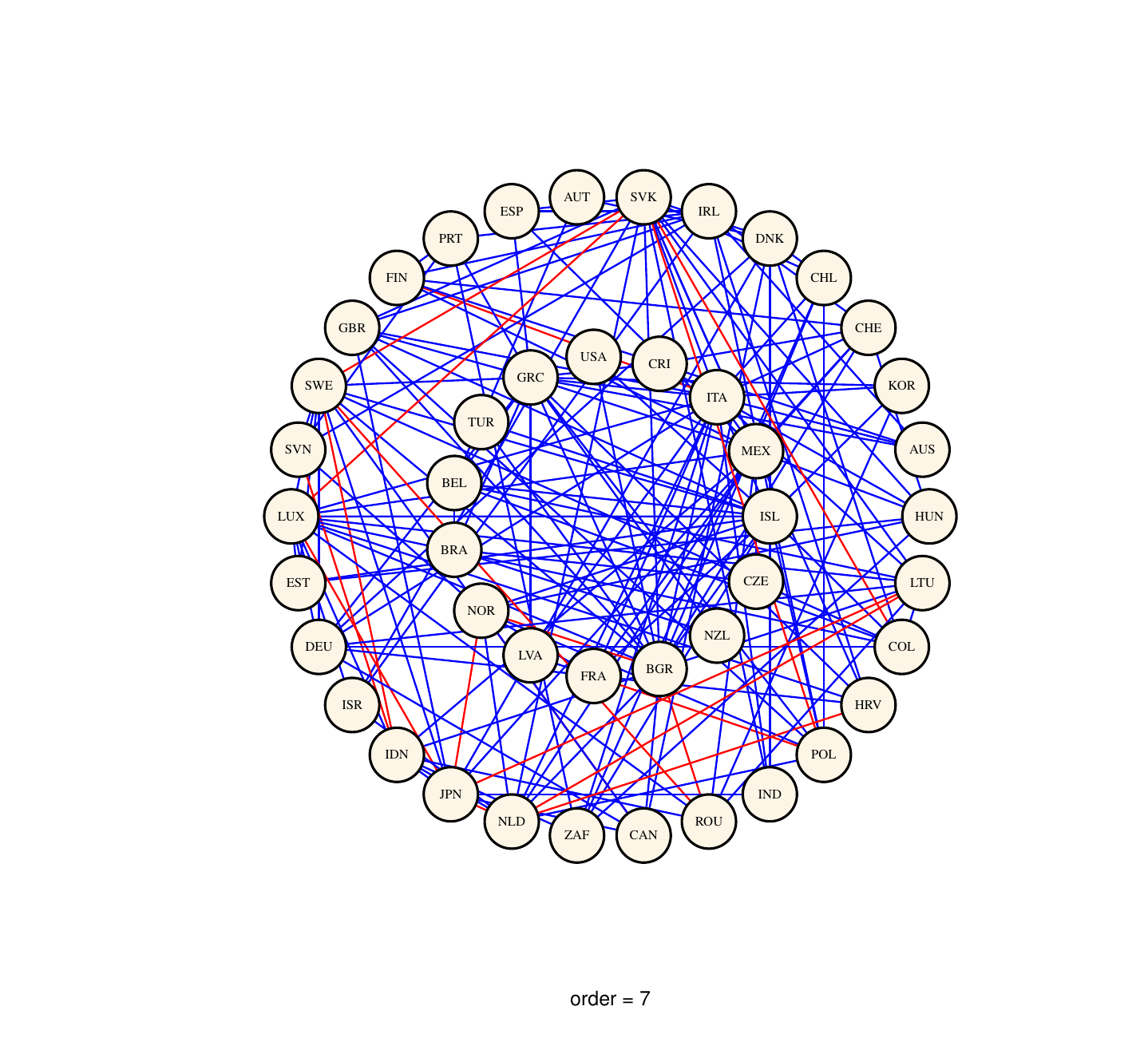}
        \caption{Overall \textit{positive shifts} in import trade relations, highlighting stronger correlations among the trading partners; using spOUTAR with order 7.}
        \label{fig:imports}
    \end{subfigure}
    \hfill
    \begin{subfigure}[b]{0.42\textwidth}
        \centering
        \includegraphics[width=\textwidth,trim=5.5cm 4.0cm 4.0cm 3.5cm,clip]{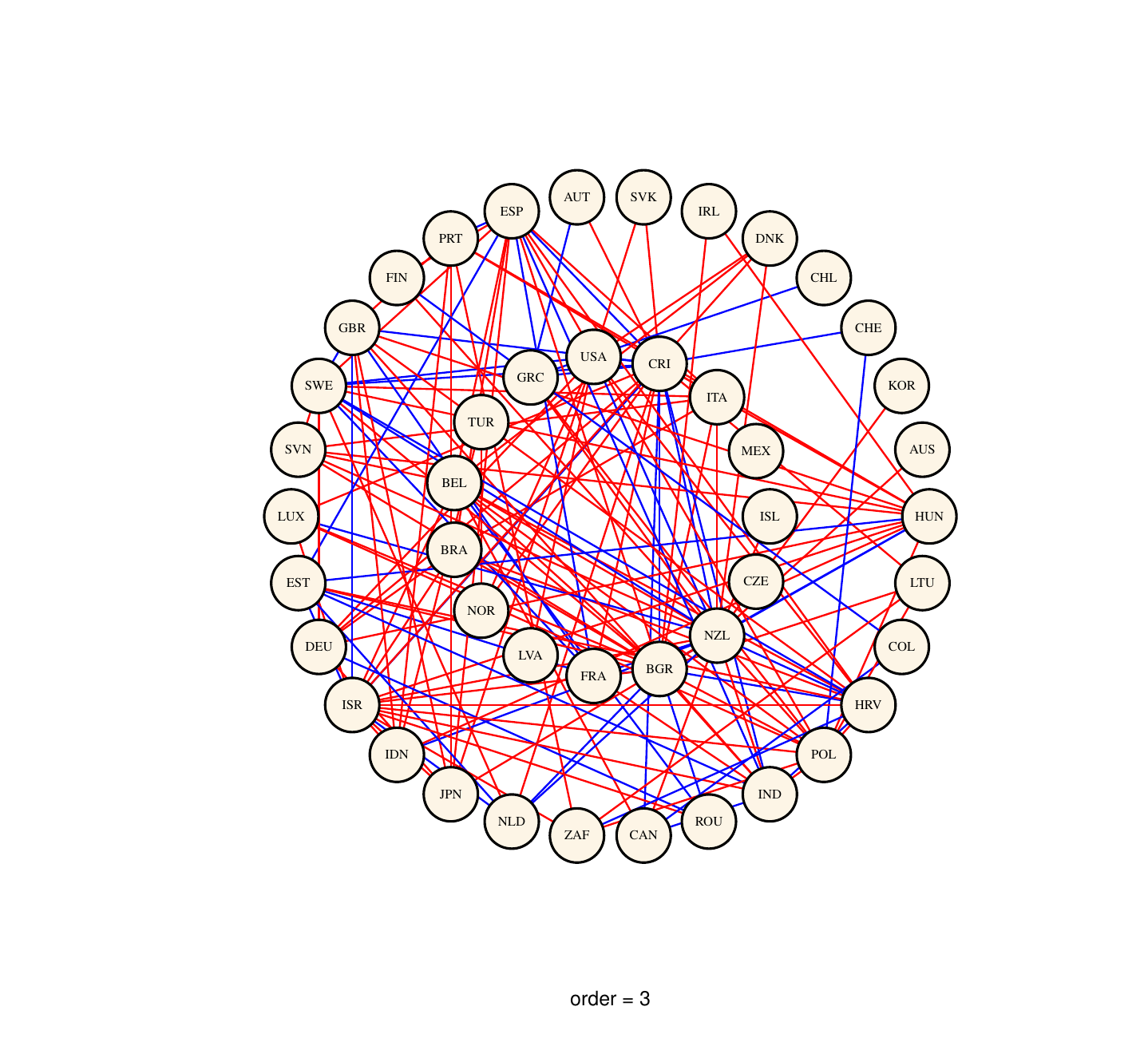}
        \caption{Overall \textit{negative shifts} in export trade relations, highlighting weaker correlations among the trading partners; using spOUTAR with order 3.}
        \label{fig:exports}
    \end{subfigure}
    \caption{Shifts in trade dynamics among the OECD nations following the Great Recession [Dec 2007 - Jun 2009]: (a) import trade volumes and (b) export trade volumes.}
    \label{fig:trade_side_by_side}
\end{figure}

\begin{figure}[h!]
    \centering
    \includegraphics[width=\textwidth,trim=2.0cm 0.8cm 2.0cm 1.6cm,clip]{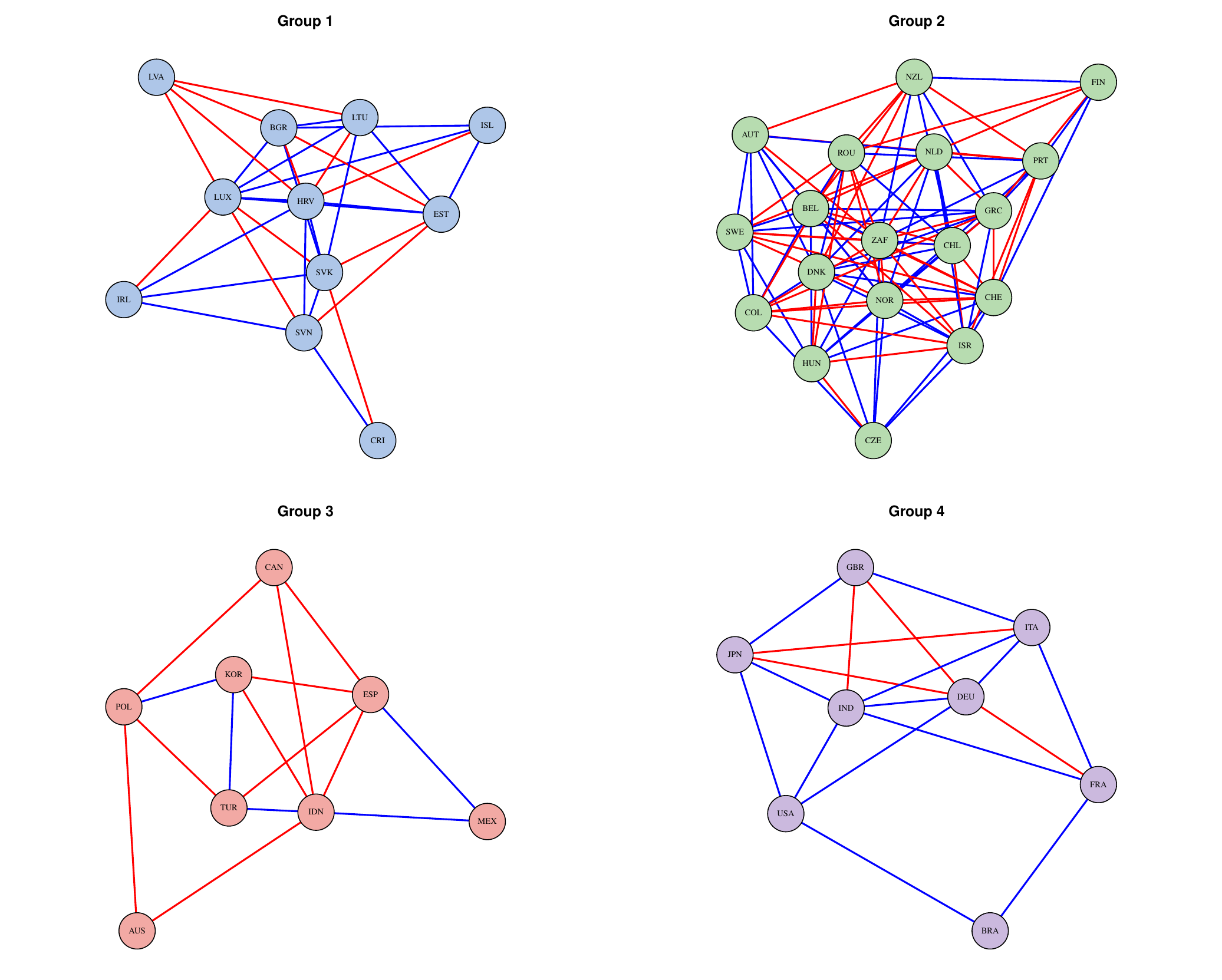}
    \caption{Group-wise classification of shifts in import trade volumes among the OECD nations following the recession as obtained from spOUTAR with order 7.}
    \label{fig:imports_grouped}
\end{figure}

\begin{figure}[h!]
    \centering
    \includegraphics[width=\textwidth,trim=2.0cm 0.8cm 2.0cm 1.6cm,clip]{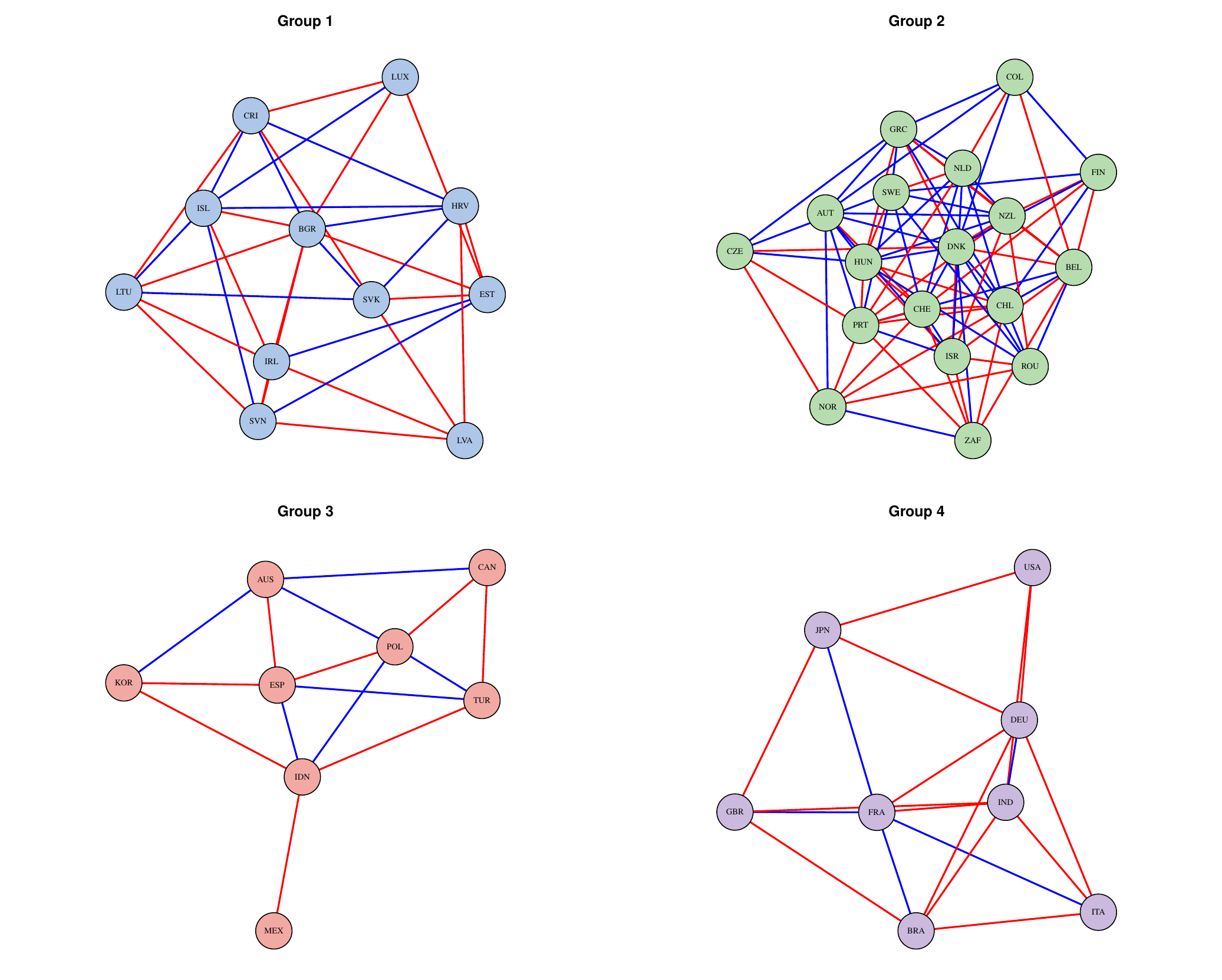}
    \caption{Group-wise classification of shifts in export trade volumes among the OECD nations following the recession as obtained from spOUTAR with order 3.}
    \label{fig:exports_grouped}
\end{figure}

The total GDP and its components exhibit a rich graphical dependency structure. We analyze the graphical patterns in import and export trade data, as shown in Figures~\ref{fig:imports} and \ref {fig:exports}, respectively. Additional graphs and the names of the OECD countries corresponding to the abbreviated identification labels in each graph's nodes are displayed in Appendix~\ref{sec:appoecd}. Furthermore, to provide a more detailed representation of structural changes, we group the OECD countries into four groups based on their GDP in the year 1997; see Table~\ref{tab:OECDgroups} in the Appendix. We show the structural changes in volumes of import trade within each group in Figure~\ref{fig:imports_grouped} and changes in export trade volumes in Figure~\ref{fig:exports_grouped}. Both import and export trade graphs reveal significant cross-country dependence. According to the World Trade Organization (WTO), global trade volumes declined by approximately 12\% during the recession. Following the crisis, global trade experienced a strong rebound, though it did not surpass pre-crisis levels until 2012, after which economic growth was slower than pre-recession trends. Our analysis indicates that the import trade graph displays an overall positive shift, whereas the export trade graph shows a predominantly negative shift. The contrasting pattern in both graphs suggests significant changes in trade relationships and policies among the $45$ countries following the crisis. Another possible reason is the uneven recovery of market share after the recession. Traditional markets, such as the United States and the European Union, experienced slower recoveries, leading to increased imports from faster-recovering emerging markets in Asia and Latin America.

\subsubsection{Quarterly macroeconomic data: Federal Reserve Bank of St. Louis}\label{sec:FRED}
In this section, we analyze quarterly macroeconomic data obtained from the Federal Reserve Bank of St. Louis \citep{mccracken2020}.
We implement the same analysis as in the previous subsection. The RMSE for one-step ahead prediction in each case is depicted in Table~\ref{tab:fred}, and the top-$10$ connections in each case are reported in Table~\ref{tab:fredcon}. spOUTAR performs better than Sparse-VAR in all cases except for the housing group. Furthermore, there was no graphical dependency among the features in the Housing group; therefore, we do not report any graphical connectivity for this group. 

\begin{table}[h!]
\centering
\caption{One-step prediction errors of Sparse-VAR and spOUTAR models for different orders of autoregressive processes for FRED-QD data.}
\resizebox{0.8\textwidth}{!}{%
\begin{tabular}{@{}llrrrrrrr@{}}
\toprule
\multirow{2}{*}{Group (USD)} &\multirow{2}{*}{Model}       & \multicolumn{7}{c}{q}                                                                                 \\ \cmidrule(l){3-9} 
                       &    & 2        & 3        & 4                 & 5                 & 6        & 7                 & 8        \\ \midrule
\multirow{2}{*}{NIPA}                & Sparse-VAR   & 0.6177 & 0.6087 & 0.7331          & 0.6496          & 0.6482 & 0.6364          & 0.6829 \\ 
                       & spOUTAR & 0.6032 & 0.6162 & 0.6578 & 0.9057          & \textbf{0.4048}  & 0.7063         & 0.4107  \\ \midrule
\multirow{2}{*}{Industrial production}                & Sparse-VAR   & 24.3526 & 23.2725 & 23.7841          & 23.9199          & 23.4976  & 23.1715          & 22.1972  \\ 
                       & spOUTAR & 1700.7240 & 20472.4000 & 345.1962          & 5235.3960 & 1010.423 & 632.9807          & \textbf{9.4566} \\ \midrule
\multirow{2}{*}{\begin{tabular}[c]{@{}l@{}}Employment and \\ unemployment\end{tabular}}                & Sparse-VAR   & 1316.0740 & 1314.5540 & 1357.4890         & 1394.3040          & 1446.8970 & 1452.6650          & 1466.6410 \\ 
                       & spOUTAR & 20.0720  & 12.8518 & 9.5959          & 18.4982 & \textbf{9.4743} & 11.9983           & 23641.34  \\ \midrule
\multirow{2}{*}{Housing}                & Sparse-VAR   & 0.1085 & 0.0830 & \textbf{0.0775} & 0.1500          & 0.0825 & 0.0920           & 0.0989  \\ 
                       & spOUTAR & 0.1176 & 0.2765 & 0.1123          & 0.1703          & 0.1249 & 0.1297          & 0.1201 \\ \midrule
\multirow{2}{*}{Prices}                & Sparse-VAR   & 41.7861 & 41.8191 & 41.8583          & 41.7098          & 41.5864 & 41.5345          & 41.4844 \\ 
                       & spOUTAR & 0.3968 & 0.1667  & 0.1804         & 0.1534          & 2.6356 & \textbf{0.1622} & 0.7116 \\ \bottomrule
\end{tabular}%
}
\label{tab:fred}
\end{table}

\begin{table}[h!]
\centering
\caption{Top 10 edges in terms of absolute vaue of conditional dependencies in FRED-QD data groups based on the sorted absolute differences in the posterior mean of $\Om_{21}$..}
\resizebox{0.8\textwidth}{!}{%
\begin{tabular}{@{}lr@{}}
\toprule
Group & Dependencies (egdes)                                                                               \\ \midrule
NIPA     & $(3, 18), (1, 13), (14, 22), (7, 16), (12, 16), (10, 12), (3, 11), (3, 5), (3, 8), (12, 22)$            \\
\begin{tabular}[c]{@{}l@{}}Industrial \\ production\end{tabular}     & $(14, 16), (14, 15), (11, 14), (4, 14), (1, 14), (10, 14), (5, 14), (2, 14), (6, 14), (7, 14)$             \\
\begin{tabular}[c]{@{}l@{}}Employment and \\ unemployment\end{tabular}     & $(26, 29), (27, 34), (25, 27), (26, 30), (27, 45), (27, 29), (34, 47), (26, 48), (26, 32), (24, 48)$ \\
Prices     & $(21, 46), (24, 38), (7, 46), (24, 41), (11, 19), (10, 20), (7, 34), (3, 23), (12, 38), (37, 42)$           \\ \bottomrule
\end{tabular}
}
\label{tab:fredcon}
\end{table}

We now examine the graphical relationships among factors in two groups: National Income and Product Accounts (NIPA) and Employment and Unemployment. The graphs for the remaining factors are presented in Appendix~\ref{sec:appfred}. Both groups exhibit a balanced pattern of positive shifts and negative shifts, indicating that the relationships between pairs of factors were reshaped significantly following the 2007--2009 recession. Notably, of the $50$ factors analyzed in the third group, the first $23$ are related to employment, while the remaining $27$ pertain to unemployment. Our analysis reveals that, in general, the relationships among unemployment-related factors were significantly affected by the recession.



\begin{figure}[h!]
    \centering
    \begin{subfigure}[b]{0.35\textwidth}
        \centering
        \includegraphics[width=\textwidth,trim=3.5cm 2cm 2.5cm 2cm,clip]{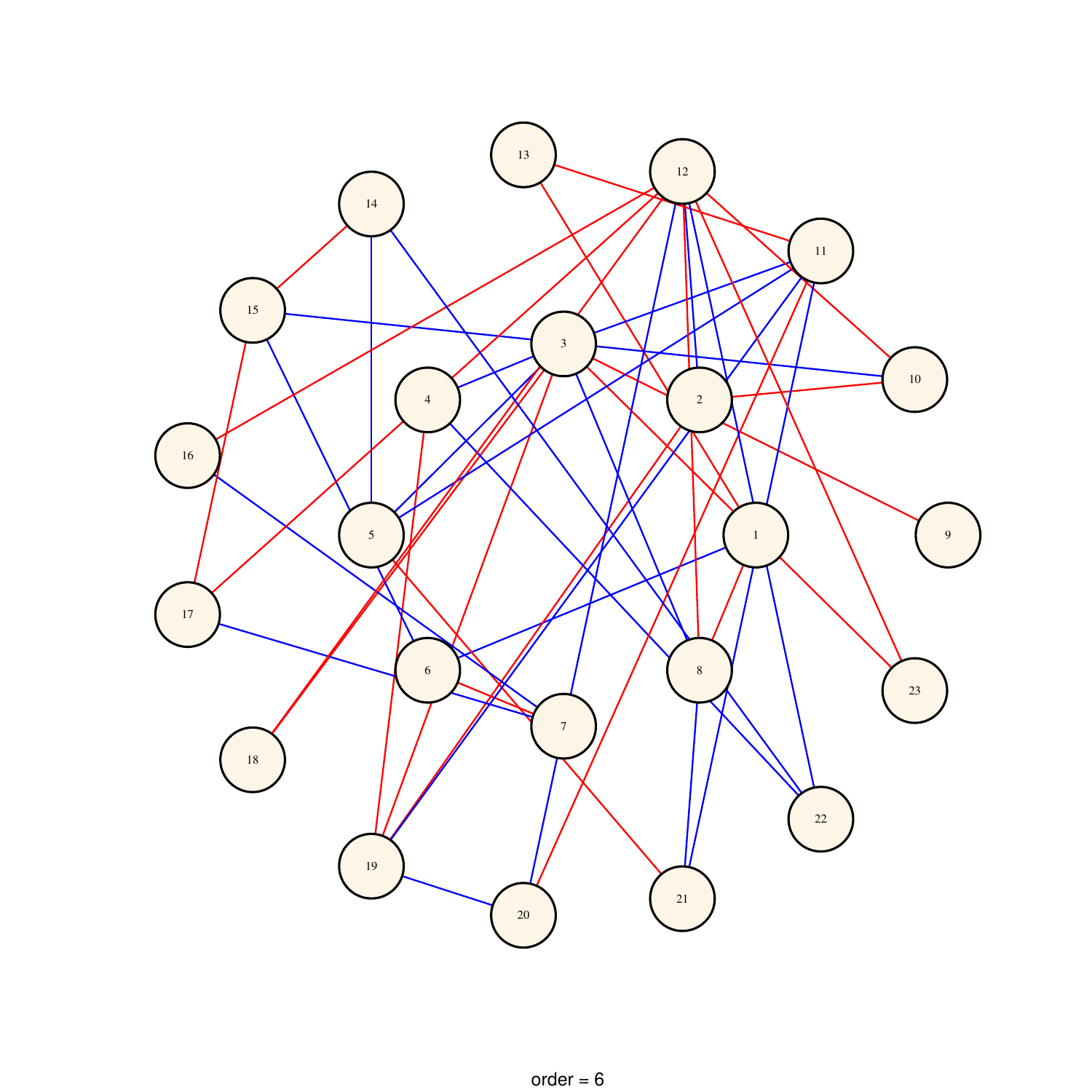}
        \caption{Changes in key NIPA indicators, highlighting the largest structural shifts.}
        \label{fig:nipa}
    \end{subfigure}
    \hfill
    \begin{subfigure}[b]{0.42\textwidth}
        \centering
        \includegraphics[width=\textwidth,trim=3.5cm 2cm 2.5cm 2cm,clip]{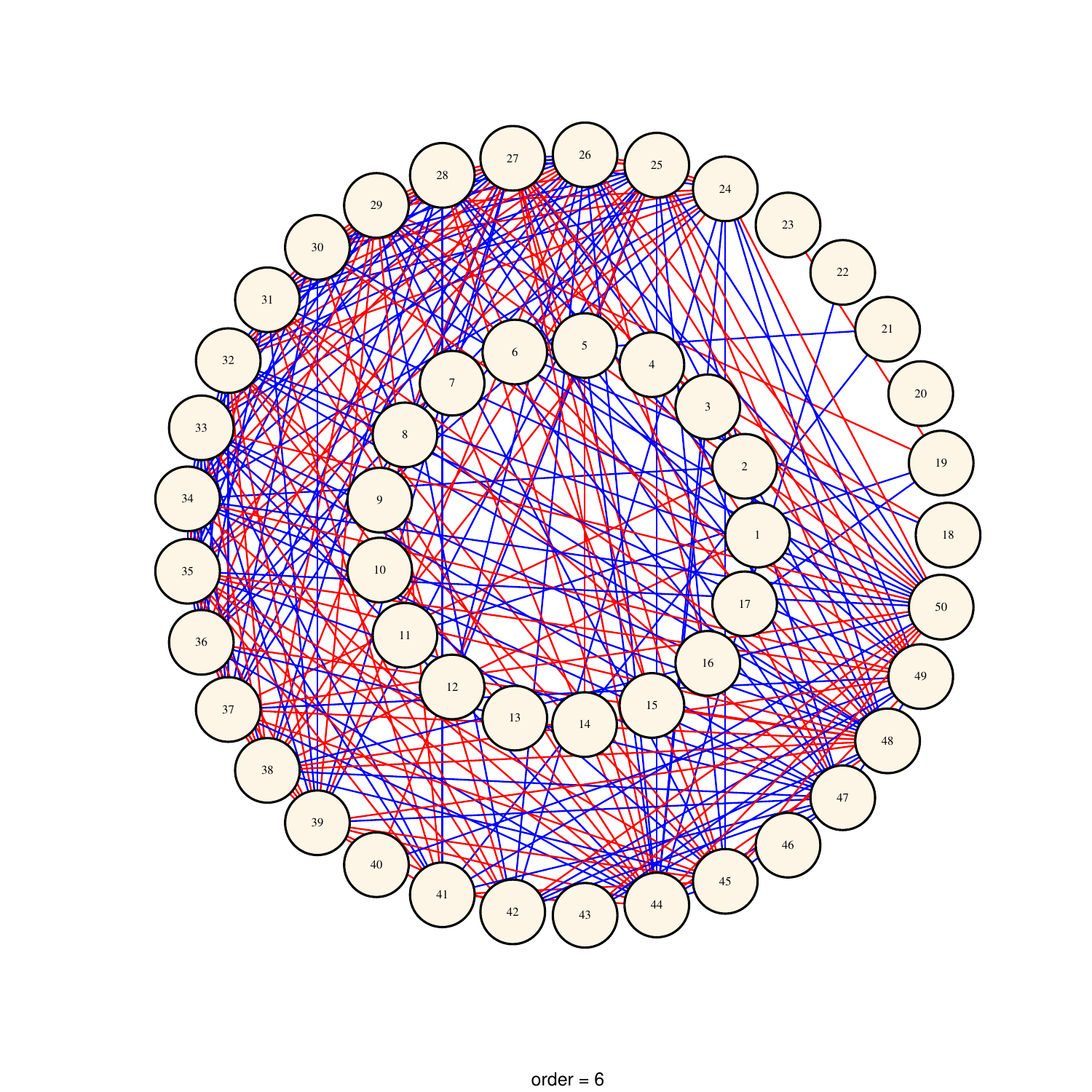}
        \caption{Changes in employment and unemployment metrics post recession, showing balanced positive and negative shifts in the labor market.}
        \label{fig:emp}
    \end{subfigure}
    \caption{Impact of the recession on the macroeconomic and labor market indicators in the US economy as revealed by spOUTAR with order 6 in: (a) NIPA attributes and (b) employment/unemployment metrics.}
    \label{fig:nipa_emp_side_by_side}
\end{figure}

\subsection{Simulated data analysis}\label{sec:sim}
The focus of this simulation study is to investigate the estimation accuracy of our proposed spOUTAR model and compare it with several potential alternatives. Data is generated under the spOUTAR framework, and we evaluate the model’s performance under both correctly specified and misspecified AR structures in the latent processes. Specifically, we generate the data as follows.

Let $n$ be the number of observations for a $p$-dimensional time series. The $p \times p$ true precision matrix $\Om_0$ used to represent the conditional dependence structure of the simulated data is drawn from a graphical Wishart distribution with degrees of freedom $10$ and a scale matrix generated from the adjacency matrices of some small-world networks following the Watts-Strogatz model. In this case, we simulate 3 small-world networks, each with 20 nodes, using the \texttt{igraph} package in $\mathrm{R}$. Then $p$ time series, each with $n$ observations, are generated using the $\mathrm{arima.sim}$ function in the \texttt{astsa R} package. For the purposes of this study, we consider AR processes subject to stationarity and causality constraints. Thus, we simulate the partial autocorrelation functions (PACFs) from $\mathrm{Unif}((-1, -0.9)\cup(0.9, 1))$ distribution and convert them to AR coefficients using the \texttt{inla.ar.PACF2phi} function of the \texttt{INLA} package. The high population PACF values are intended to induce strong temporal dependence in the simulated data, thereby making the numerical experiments more challenging. The autocorrelation functions are generated using \texttt{inla.ar.PACF2acf} and the innovation variances of the AR models are specified such that their marginal variances are 1 as mentioned in Section \ref{sec:model}. Given the true order $q$, we construct the $p \times n$ matrix $\Z$. The same value of $q$ is used to fit the spOUTAR model, assuming the AR process orders are known. The final data is obtained as $\Y = \s_0^{1/2}\Z$ where $\s_0^{1/2}$ is the square root of the inverse of $\Om_0$.

\subsection{Correctly specified AR orders}

In this case, we assume that the true order of the latent AR processes underlying the data generation is known and use this order to fit a spOUTAR model. We consider three levels of sparsity for $\Om_0$, namely, $75\%$, $90\%$, and $95\%$. For each level of sparsity, we choose the dimension of the data, sample size, and the order of the AR processes from the following grid: $p\in\{30,60,90\}$, $n\in\{50,100,150\}$, and $q\in\{2,5,8,10\}$. We compare the performance of our method with that of GGM and the Gaussian copula-based graphical model (GCGM; see \cite{mohammadi2017}). The average root mean squared error: $\mathrm{RMSE} = \|\bar{\Om} - \Om_0\|_{\mathrm{F}}/p$ across all experiments for $90\%$ sparsity case is displayed in Table~\ref{tab:sim2}, and the results for the other two cases can be found in Appendix~\ref{sec:appsim}.

\begin{table}[h!]
\centering
\resizebox{0.70\textwidth}{!}{
\begin{tabular}{@{}llcrrrllcrrr@{}}
\toprule
\text{$p$} & \text{$n$} & \text{$q$} & \text{GGM} & \text{GCGM} & \text{spOUTAR} & \text{$p$} & \text{$n$} & \text{$q$} & \text{GGM} & \text{GCGM} & \text{spOUTAR}\\ \midrule
30 & 50  & 2 & 6.0377 & \textbf{2.6564} & 3.4017 & 30 & 150 & 8 & 26.0452 & 5.7815 & \textbf{3.7011} \\
30 & 100 & 2 & 10.4520 & \textbf{2.8323} & 3.7694 & 30 & 200 & 8 & 33.5753 & 6.5204 & \textbf{3.7361} \\
30 & 150 & 2 & 14.8057 & \textbf{2.7953} & 3.7674 & 30 & 250 & 8 & 41.6565 & 7.4157 & \textbf{3.8038} \\
60 & 50  & 2 & 5.3626 & \textbf{1.9444} & 2.2999 & 60 & 150 & 8 & 19.8279 & 5.1945 & \textbf{2.7491} \\
60 & 100 & 2 & 8.8180 & \textbf{1.9729} & 2.6945 & 60 & 200 & 8 & 25.6120 & 5.6582 & \textbf{2.6156} \\
60 & 150 & 2 & 12.0416 & \textbf{2.0238} & 2.6933 & 60 & 250 & 8 & 31.3362 & 6.7175 & \textbf{2.6915} \\
90 & 50  & 2 & 5.2464 & 1.9480 & \textbf{1.8120} & 90 & 150 & 8 & 17.7364 & 5.1277 & \textbf{2.2704} \\
90 & 100 & 2 & 7.9137 & \textbf{1.8539} & 2.2087 & 90 & 200 & 8 & 21.9630 & 5.8949 & \textbf{2.2408} \\
90 & 150 & 2 & 9.6735 & \textbf{1.6800} & 2.3388 & 90 & 250 & 8 & 26.9005 & 5.7060 & \textbf{2.3510} \\
\midrule
30 & 50  & 5 & 8.4568 & \textbf{2.8745} & 3.5941 & 30 & 150 & 10 & 27.1173 & 7.3088 & \textbf{3.7588} \\
30 & 100 & 5 & 15.6330 & \textbf{3.3551} & 3.5906 & 30 & 200 & 10 & 35.6630 & 7.9388 & \textbf{3.8178} \\
30 & 150 & 5 & 22.1923 & 3.7786 & \textbf{3.6933} & 30 & 250 & 10 & 43.9372 & 10.1274 & \textbf{3.7867} \\
60 & 50  & 5 & 7.1477 & 2.6807 & \textbf{2.5512} & 60 & 150 & 10 & 20.7453 & 6.0189 & \textbf{2.6153} \\
60 & 100 & 5 & 12.2996 & 3.1222 & \textbf{2.7351} & 60 & 200 & 10 & 26.9059 & 7.6256 & \textbf{2.7214} \\
60 & 150 & 5 & 20.1701 & 3.7116 & \textbf{3.2150} & 60 & 250 & 10 & 32.9758 & 7.0657 & \textbf{2.5505} \\
90 & 50  & 5 & 6.6208 & 2.9410 & \textbf{1.9337} & 90 & 150 & 10 & 18.2535 & 6.1096 & \textbf{2.1994} \\
90 & 100 & 5 & 10.8862 & 2.5793 & \textbf{2.2669} & 90 & 200 & 10 & 23.2690 & 6.1435 & \textbf{2.2679} \\
90 & 150 & 5 & 15.0431 & 2.9079 & \textbf{2.3402} & 90 & 250 & 10 & 28.6306 & 7.1274 & \textbf{2.1516} \\
\bottomrule
\end{tabular}
}
\caption{Correctly-specified AR case: Comparison of average RMSE value in the precision matrix estimation where the true precision is $90\%$ sparse, following a small world networks from the Watts-Strogatz model for different choices of dimension $p$, sample size $n$, and AR order $q$.}
\label{tab:sim2}
\end{table}

The GGM performs the worst in all three cases. However, for lower values of $q$, the GCGM performs best in most scenarios. This may be attributed to the stronger inductive biases of the GCGM model in relatively low-dimensional spaces associated with lower-order AR processes, as well as the fact that although the underlying temporal dependence is strong, in this case its effects decay relatively quickly over time. Furthermore, across different series, copulas are well-known for separating the modeling of marginal distributions from the dependence structure, enabling independent modeling of each variable's distribution while capturing their joint behavior. A less sparse underlying graph further enhances its strong performance. However, as the order of the AR process increases or the underlying graphical structure becomes sparser, our method achieves the best performance.

\subsection{Misspecified AR orders}

In practice, it is difficult to know the true order of the AR processes. Further, it may happen that all the autoregressions do not exhibit the same degree of dependency on their past observations. Thus, it is important to assess the credibility of spOUTAR in such situations. We set up a simulation study in which the underlying data are generated from a mixture of ARs with different lag orders. Let $m_1, m_2$, and $m_3$ denote the number of simulated AR processes of orders 2, 5, and 8. We vary the numbers $m_1, m_2$, and $m_3$ in such a way that every order dominates the other two at some time. The sparsity level of $\Om_0$ is set at $90\%$. While training the spOUTAR model, we set the AR order $ q$ for all univariate series to the same value and vary it from 2 to 8.

Table \ref{tab:mixar} illustrates the results obtained from the simulation. It is clear that the proposed method delivers the best performance across all the cases. As the temporal dependencies among observations in each time series become more complex, the GCGM performs poorly, likely because it cannot capture higher-order temporal dependencies. 


\begin{table}[h!]
\centering
\resizebox{\textwidth}{!}{
\begin{tabular}{@{}clclcccclclcccclclccc@{}}
\toprule
  \text{$q$} &\text{$m$} &\text{GGM} &\text{GCGM} &\text{spOUTAR} &\text{$m$} &\text{GGM} &\text{GCGM} &\text{spOUTAR} &\text{$m$} &\text{GGM} &\text{GCGM} &\text{spOUTAR}
  \\ \midrule
2 & (30, 15, 15) & 11.3004 & 2.9574 & \textbf{2.6534} &
(15, 30, 15) & 11.9860 & 3.4435 & \textbf{2.6379} &
(15, 15, 30) & 12.2606 & 3.8291 & \textbf{2.6708} \\

3 & (30, 15, 15) & 11.1240 & 3.3355 & \textbf{2.7101} &
(15, 30, 15) & 12.1674 & 3.5209 & \textbf{2.7009} &
(15, 15, 30) & 12.2681 & 3.9956 & \textbf{2.6240} \\

4 & (30, 15, 15) & 11.2707 & 3.4013 & \textbf{2.6024} &
(15, 30, 15) & 12.0825 & 3.3375 & \textbf{2.7458} &
(15, 15, 30) & 12.3202 & 3.6375 & \textbf{2.6937} \\

5 & (30, 15, 15) & 11.1912 & 3.5026 & \textbf{2.9191} &
(15, 30, 15) & 11.8330 & 3.2521 & \textbf{2.7774} &
(15, 15, 30) & 12.2782 & 3.8517 & \textbf{2.6693} \\

6 & (30, 15, 15) & 11.3114 & 3.2632 & \textbf{2.6638} &
(15, 30, 15) & 11.8977 & 3.3793 & \textbf{2.8290} &
(15, 15, 30) & 12.4350 & 3.6204 & \textbf{2.6603} \\

7 & (30, 15, 15) & 11.3305 & 3.1309 & \textbf{2.7027} &
(15, 30, 15) & 12.1298 & 3.5228 & \textbf{3.2474} &
(15, 15, 30) & 12.2040 & 3.6489 & \textbf{2.6879} \\

8 & (30, 15, 15) & 11.2302 & 3.1590 & \textbf{2.6862} &
(15, 30, 15) & 11.9044 & 3.2057 & \textbf{2.7256} &
(15, 15, 30) & 12.3466 & 3.7656 & \textbf{2.7221} \\
\bottomrule
\end{tabular}
}
\caption{Misspecified AR case: Comparing average RMSE of the precision matrix estimates, where the latent AR processes are generated by combining $m_1$ time series of order 2, $m_2$ time series of order 5, and $m_3$ time series of order 8; denoted by the tuple $m:=(m_1,m_2,m_3)$. The sparsity level of the true precision is set at $90\%$, and spOUTAR is fitted, setting $q$ as the order for all the univariate series. 
}
\label{tab:mixar}
\end{table}

\section{Discussion}\label{sec:diss}
This article presents a novel model for estimating changes in the underlying conditional dependency structure in multivariate time series data. Based on the simulation studies, $\OUT$ outperforms other competing methods, particularly when the data exhibit high degree of temporal correlation among observations or the underlying graphical structure is very sparse. Furthermore, the proposed model demonstrates superior predictive performance across all real-data analyses, effectively estimating significant changes in relationships among macroeconomic factors in the US and among GDP and related components in OECD countries following the Great Recession (2007–2009). As an extension to this work, it would be interesting to see how the $\OUT$ model performs with univariate latent nonlinear autoregressive processes.

\section*{Funding}
The authors would like to thank the National Science Foundation collaborative research grants DMS-2210280 (Shuvrarghya Ghosh, Subhashis Ghosal) / 2210281 (Anindya Roy) / 2210282 (Arkaprava Roy).

\bibliography{references}

\appendix
\section{Appendix}

\subsection{Computation of the gradients}
In this section, we derive the gradients of the joint log-likelihood function of $p$ many AR($q$) processes, each having $n$ observations, with respect to $\IL$ and $\D$. Without loss of generality, we set $\Om_1=\Om_2=\Om$ for simpler computations. In case of two precision matrices, we set $\Z=\Z_1$ in the computations detailed in Section~\ref{app:z}. Subsequently, utilize the result in Section~\ref{app:l} to obtain the gradient with respect to $\IL_1$. Likewise, the gradient of $\IL_2$ can be computed taking $\Z=\Z_2$. The notation $\nabla_\mathbf{X}f$ denotes the vector differential operator that computes the gradient of a function $f:\mathbb{R}^d\to\mathbb{R}$ with respect to $\vect(\mathbf{X})$. Lastly, we define $\vect^{-1}: \mathbb{R}^{mn} \to \mathbb{R}^{m \times n}$ as an inverse vectorization map. 

\subsubsection{Gradient with respect to $\Z$}\label{app:z}

The joint log-likelihood function of all latent AR processes of order $q$ is expressed as
\[\log\pi(\Psi, \boldsymbol{\sigma}|\Z) = -\sum_{i=1}^p\frac{1}{2\sigma_j^2} \sum_{t=q+1}^n \big(Z_{j,t} - \sum_{k=1}^q \varphi_{j, k} Z_{j,t-k}\big)^2.\]
The derivative of this function with respect to $\Z$ is a matrix of dimensions $p \times n$ whose $(j, t)$th entry is computed as follows:
\[
\left(\nabla_{\Z}\log\pi(\Psi, \boldsymbol{\sigma}\mid \Z)\right)_{j,t}
=
\begin{cases}
-\dfrac{1}{\sigma_j^2}
(
Z_{j,t} - \displaystyle\sum_{k=1}^q \varphi_{j,k} Z_{j,t-k}
)
\\[0.7em]
\quad
+
\displaystyle\sum_{j=1}^{\min(N-t,\,q)}
\dfrac{1}{\sigma_j^2}
(
Z_{j,t+s} - \sum_{k=1}^q \varphi_{j,k} Z_{j,t+s-k}
)
\varphi_{j,k},
& t > q,
\\[1.2em]
\dfrac{1}{\sigma_j^2}
(
Z_{j,q+1} - \displaystyle\sum_{k=1}^q \varphi_{j,k} Z_{j,q+1-k}
)
\varphi_{j,k},
& t \in [q].
\end{cases}
\]



\subsubsection{Gradient with respect to L}\label{app:l}
We know $\Om^{1/2} = \D(\I-\IL)$ such that $\Z = \U^{\mathrm{T}}\D(\I-\IL)\Y$. Vectorizing both sides of the latter equation, we have $$\vect(\Z) = \vect(\U^{\mathrm{T}}\D\Y) - \vect(\U^{\mathrm{T}}\D\IL\Y)\mbox{ and }\vect(\U^{\mathrm{T}}\D\IL\Y) = \left[\Y^{\mathrm{T}} \otimes \U^{\mathrm{T}}\D\right]\vect(\IL).$$ 
Since $\nabla_{\IL}\vect(\Z) = -(\Y^{\mathrm{T}} \otimes \U^{\mathrm{T}}\D)$, applying the chain rule of differentiation, we obtain 
\[ \nabla_{\IL}\log\pi(\Psi, \boldsymbol{\sigma}|\Z) = - \vect^{-1}\Big[ \left[\Y \otimes \D\U\right] \times \vect\big( \nabla_{\Z}\log\pi(\Psi, \boldsymbol{\sigma}|\Z) \big) \Big]. \]

\subsubsection{Gradient with respect to log(D)}\label{app:d}
We can write $\vect(\Z) = \left[\Y^{\mathrm{T}}(\I - \IL)^{\mathrm{T}} \otimes \U^{\mathrm{T}} \right]\vect(\D)$. Differentiating both sides with respect to $\D$, we have 
$$\nabla_{\D}\vect(\Z) = \Y^{\mathrm{T}}(\I - \IL)^{\mathrm{T}} \otimes \U^{\mathrm{T}} \mbox{ and }   \nabla_{\log\D} \vect(\D) = \vect(\D),$$
where $\otimes$ denotes the Kronecker product. By the chain-rule, we obtain
\[ \nabla_{\log\D}\log\pi(\Psi, \boldsymbol{\sigma}|\Z) = \vect^{-1}\Big[\big((\I - \IL)\Y \otimes \U \big) \vect\big( \nabla_{\Z}\log\pi(\Psi, \boldsymbol{\sigma}|\Z) \big)\Big] \odot  \D. \] 
 Note that in the Langevin Monte Carlo steps, we evaluate the negative log-likelihood function and its gradient.

\subsection{Supplementary results: FRED-QD data analysis}\label{sec:appfred}
\begin{figure}[h!]
    \centering
    \begin{subfigure}[b]{0.35\textwidth}
        \centering
        \includegraphics[width=\textwidth,trim=3.5cm 2cm 2.5cm 2cm,clip]{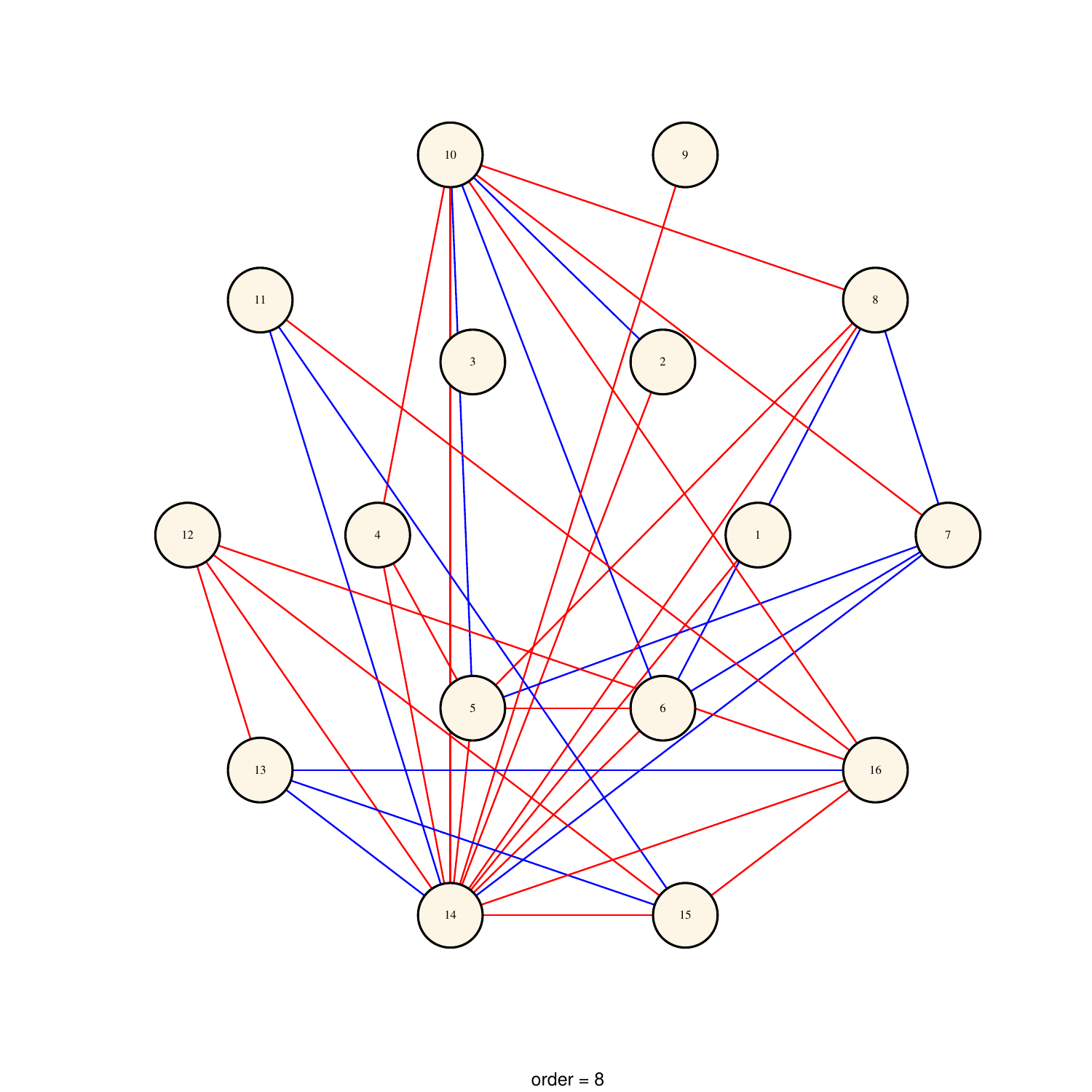}
        \caption{Shifts in industrial production indicators, showing sector-specific output changes, as revealed by spOUTAR with order 8.}
        \label{fig:indus}
    \end{subfigure}
    \hspace{35pt}
    \begin{subfigure}[b]{0.40\textwidth}
        \centering
        \includegraphics[width=\textwidth,trim=3.5cm 2cm 2.5cm 2cm,clip]{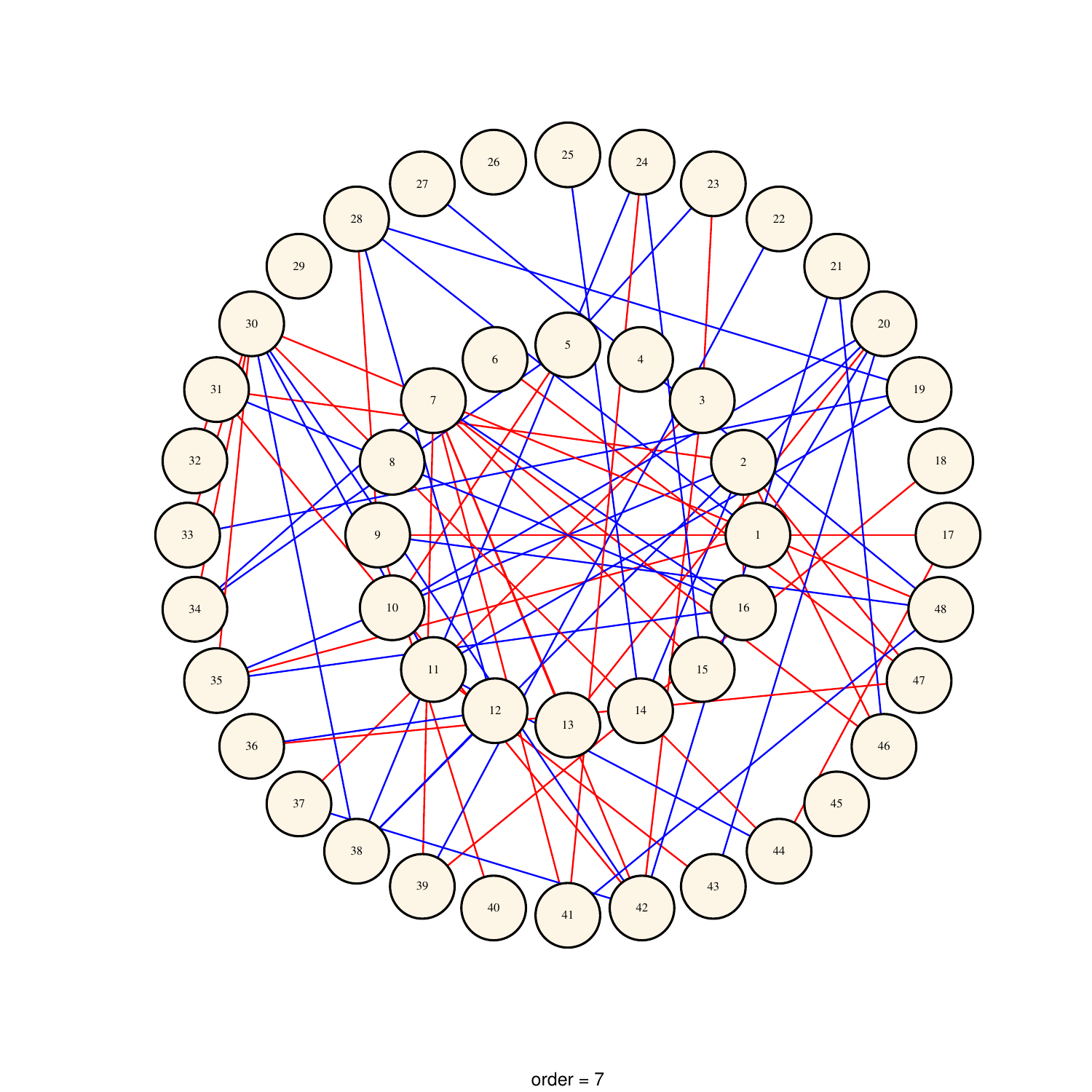}
        \caption{Shifts in price-related attributes (inflation, CPI components), as revealed by spOUTAR with order 7.}
        \label{fig:price}
    \end{subfigure}
    \caption{Economic impact of the 2009 recession on: (a) industrial output, (b) price changes across sectors.}
    \label{fig:indus_price_side_by_side}
\end{figure}

The graphs for Industry attributes and Prices attributes are presented in Figure~\ref{fig:indus} and Figure~\ref{fig:price}, respectively.

\subsection{Proof of the main theorem}
\label{sec:theorem proofs}

We apply the following general rate theorem given in Theorem~1 of \cite{Roy2024}. 

\begin{theorem}[General pseudo-posterior contraction rate]
\label{thm:pseudoposterior general}
Let $q_\theta^{(T)}$ be a statistical model for a sequence of observations $\bY^{(T)}$ indexed by $\theta\in \Theta_T$, $T=1,2,\ldots$, and let $p_0^{(T)}$ be the true distribution of $\bY^{(T)}$. Assume that for some fixed $b>1$ and $\theta_0=\theta_{0,T}\in \Theta_T$,   
\begin{align}
    \label{eq:moment control}
   \int (p_0^{(T)}/q_{\theta_0}^{(T)})^b q_{\theta_0}^{(T)} \lesssim G_T<\infty
\end{align}
for some sequence $G_T>0$. 
Let $d_T$ be a semimetric on $\theta_T$. Let $\Pi$ stand for a prior distribution for $\theta\in \Theta_T$. Suppose that there exist subsets a sequence $\epsilon_T\to 0$, $T\epsilon_T^2\to \infty$, sets $\Theta_T^*\subset \Theta_T$, a covering $\{\Theta_{T,j}^*: j=1,\ldots,N_T\}$ of $\Theta_T^*$, test functions $\phi_{T,j}$ such that the following conditions hold.
\begin{enumerate}
    \item [{\rm (i)}] $-\log \Pi ( B(\theta_0,\epsilon_T)) \lesssim T\epsilon_T^2$, where $B(\theta_0,\epsilon_T)= \{\theta: \int p_{0}^{(T)} \log (q_{\theta_0}^{(T)}/q_\theta^{(T)}) \le T\epsilon_T^2 \}$; 
     \item [{\rm (ii)}] $\int \phi_{T,j} q_{\theta_0}^{(T)} \le e^{-K_0 T\epsilon_T^2}$ and $\sup \{ \int (1-\phi_{T,j}) q_\theta^{(T)}: d_T(\theta_0,\theta)>\epsilon_T, \theta\in \Theta_{T,j}^*\} \le e^{-K_0 T\epsilon_T^2}$ for some $K_0>0$;
     \item [{\rm (iii)}]  $\log N_T \lesssim T\epsilon_T^2$;
     \item [{\rm (iv)}] $e^{cT\epsilon_T^2} \Pi (\Theta_T\setminus \Theta_T^*)/  \Pi (B(\theta_0,\epsilon_T))\to 0$ for any fixed $c>0$;
     \item [{\rm (v)}] $\log G_T \lesssim T\epsilon_T^2$.
\end{enumerate}
Then for every $m_T\to \infty$, $\Pi (d_T( \theta_0,\theta)> m_T \epsilon_T|\bY^{(T)})\to 0$ in $P_{0}^{(T)}$-probability. 
\end{theorem}

\begin{proof}[Proof of Theorem~\ref{thm:main}]
We only give a sketch of the proof. First, we verify Properties (i)--(iii) in the proof of Lemma 6 of \cite{Roy2024}. The first two properties are the same, as we use the same priors on $\bOmega$ and $\bA$ as they did, and we assume the same conditions on these parameters. For Property (iii), we show that 
\begin{align} 
\label{eq:prior concentration}
-\log \Pi (\max\{ \|f_j-f_{j0}\|_\infty: j\in [p]\}  \le \epsilon) \lesssim \log (1/\epsilon),
\end{align}
where $f_j$ stands for the spectral density of the $j$th latent time series and $f_{j0}$ its true value, $j\in [p]$. This suffices as $f_j$, $j\in[p]$, are bounded below (and above) as argued in the proof of Lemma 6 of \cite{Roy2024}, and the boundedness property holds by the assumption that the autoregressive parameters stay within a compact interval. 

The spectral density $f_j$ of the $j$th autoregressive process is given by 
\begin{align}
  \label{eq:AR spectral}
  f_j(\omega) = \frac{\sigma_j^2}{2\pi}\times\bigg|{1+\sum_{k=1}^{q}\varphi_{j,k} e^{- i k  \omega}}\bigg|^{-2}.
\end{align}
It is immediate that the function $f_j$ is Lipschitz continuous of order $1$ in the parameter $(\varphi_{j,k}:\, k\in[q], \, \sigma_j)$ at its true value, $j\in [p]$. Since the correspondence between the PACF parameters and the AR coefficients is continuously invertible and smooth, it follows that the event in \eqref{eq:prior concentration} contains an event of the form $\{ \max_{j,k} |\rho_{j,k}-\rho_{0;j,k}|\le c\epsilon\}$ for some $c>0$, where $\rho_{0;j,k}$ is the true value of the PACF parameter $\rho_{j,k}$. Hence \eqref{eq:prior concentration} follows from the continuity and positivity of the prior densities of the PACF parameters at their true values assumed in (A5). 

To apply the rest of the arguments used in the proof of Theorem~4 of \cite{Roy2024}, it remains to check that the spectral densities are bounded and bounded away from zero. Since the values of the AR parameters lie in a compact set as a consequence of the homeomorphism connecting the PACF parameters and the AR coefficients, it is immediate from the expression in \eqref{eq:AR spectral} that the spectral densities are bounded below. Further, Condition (A3) implies that the AR parameters also stay away from the set where the expressions $\zeta_j(\omega|\varphi_{j,k}: k\in [q])=1+\sum_{k=1}^{q}\varphi_{j,k} e^{- i k  \omega} $, $j\in [p]$, vanish. Hence, the required condition holds. 

Thus, the pre-rate $\bar{\epsilon}_T$ corresponding to the spectral densities is given by the relation $T\bar\epsilon_T^2 \asymp p\log (1/\bar\epsilon_T)$, or $\bar \epsilon_T=\sqrt{p(\log T)/T}$. 
Similarly, the pre-rate $\bar{\epsilon}_T$ corresponding to the rotation $\bU$ and the precision matrix $\bOmega$ is given by $T\bar\epsilon_T^2 \asymp (p+s)\log (1/\bar\epsilon_T)$, or $\bar \epsilon_T=\sqrt{(p+s)(\log T)/T}$. 
We also note that the covering numbers can be bounded as in the proof of Lemma~8 of \cite{Roy2024}. The only difference in the argument is that the difference in spectral densities is bounded by a multiple of the sum of the differences of the PACF parameters, rather than those of the B-spline coefficients. The former live in a $pq$-dimensional space instead of a $pJ$-dimensional space, where $J\asymp \epsilon^{-\alpha}$ for $\alpha$ the smoothness index of the spectral densities. Thus, the entropy bound is $(p+s)\log (1/\epsilon)$, which is better than the corresponding value in \cite{Roy2024} even after their dimension reduction assumption in their Equation (3.7). 
Hence, in view of the compactness of the domains of the parameters and the range of the spectral densities, we do not need to use a sieve, as a fixed compact set has the full prior probability. Thus Part (d) of Theorem~4 of \cite{Roy2024} applies, with $\epsilon_T$ there replaced by 
$\sqrt{(p+s)(\log p)/T}$, and the polynomial factor in the rate characterization in Theorem~4 of \cite{Roy2024} does not appear. This completes the proof. 
\end{proof}

\subsection{Supplementary results: OECD data analysis}\label{sec:appoecd}

\begin{table}[h!]
\centering
\resizebox{0.95\textwidth}{!}{
\begin{tabular}{r l r l r l r l r l}
\toprule
\textbf{ID} & \textbf{Country} & \textbf{ID} & \textbf{Country} & \textbf{ID} & \textbf{Country} & \textbf{ID} & \textbf{Country} & \textbf{ID} & \textbf{Country} \\
\midrule
ISL & Iceland & NOR & Norway & CHE & Switzerland & GBR & United Kingdom & NLD & Netherlands \\
MEX & Mexico & LVA & Latvia & CHL & Chile & SWE & Sweden & ZAF & South Africa \\
ITA & Italy & FRA & France & DNK & Denmark & SVN & Slovenia & CAN & Canada \\
CRI & Costa Rica & BGR & Bulgaria & IRL & Ireland & LUX & Luxembourg & ROU & Romania \\
USA & United States & NZL & New Zealand & SVK & Slovakia & EST & Estonia & IND & India \\
GRC & Greece & CZE & Czech Republic & AUT & Austria & DEU & Germany & POL & Poland \\
TUR & Turkey & HUN & Hungary & ESP & Spain & ISR & Israel & HRV & Croatia \\
BEL & Belgium & AUS & Australia & PRT & Portugal & IDN & Indonesia & COL & Colombia \\
BRA & Brazil & KOR & South Korea & FIN & Finland & JPN & Japan & LTU & Lithuania \\
\bottomrule

\end{tabular}}
\caption{List of countries analyzed in OECD data with their ISO 3166 country codes.}
\label{tab:count_names}
\end{table}

\begin{table}[h!]
\centering
\resizebox{0.75\textwidth}{!}{%
\begin{tabular}{@{}rrrrr@{}}
\toprule
\multicolumn{1}{c}{Group 1} & \multicolumn{2}{c}{Group 2} & \multicolumn{1}{c}{Group 3} & \multicolumn{1}{c}{Group 4} \\ \midrule
Iceland         & New Zealand & Portugal     & Poland    & Germany        \\
Estonia         & Israel      & Austria      & Australia & Brazil         \\
Latvia          & Denmark     & Netherlands  & Turkiye   & India          \\
Luxembourg      & Hungary     & Colombia     & Korea     & Russia         \\
Costa Rica      & Finland     & Switzerland  & Canada    & United Kingdom \\
Lithuania       & Sweden      & Belgium      & Spain     & France         \\
Slovenia        & Chile       & South Africa & Indonesia & Italy          \\
Ireland         & Norway      &              & Mexico    & Japan          \\
Croatia         & Czechia     &              &           & United States  \\
Slovak Republic & Greece      &              &           &                \\
Bulgaria        & Romania     &              &           &               
\end{tabular}%
}
\caption{Classification of Countries by their GDP in 1997 (US Dollars).}
\label{tab:OECDgroups}
\end{table}

\begin{figure}[h!]
    \centering
    \begin{subfigure}[b]{0.40\textwidth}
        \centering
        \includegraphics[width=\textwidth,trim=4.4cm 3.2cm 3.3cm 2.5cm,clip]{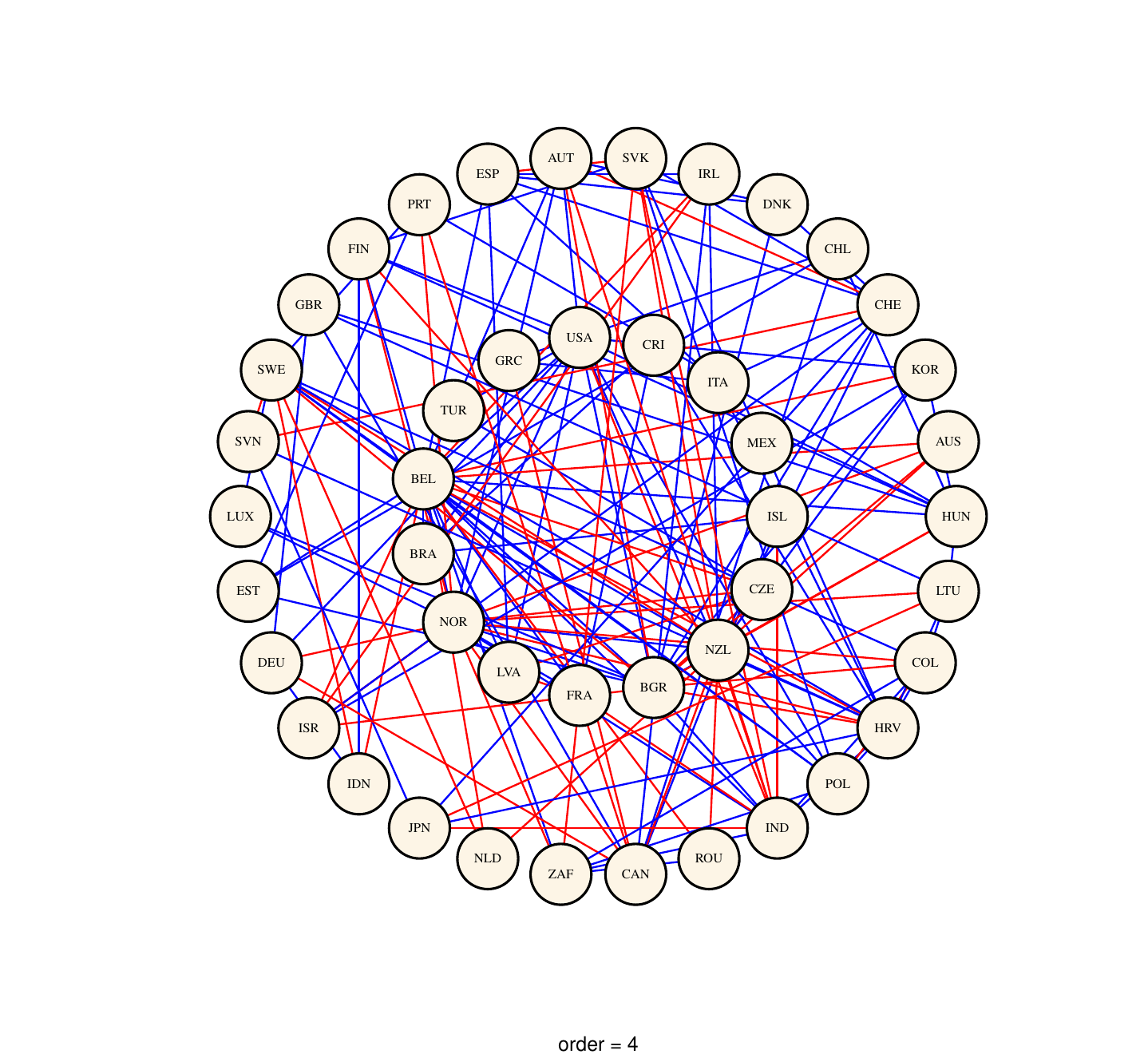}
        \caption{Changes in government expenditure indicators, highlighting fiscal shifts; using spOUTAR with order 4.}
        \label{fig:govexp}
    \end{subfigure}
    \hfill
    \begin{subfigure}[b]{0.40\textwidth}
        \centering
        \includegraphics[width=\textwidth,trim=4.4cm 3.2cm 3.3cm 2.5cm,clip]{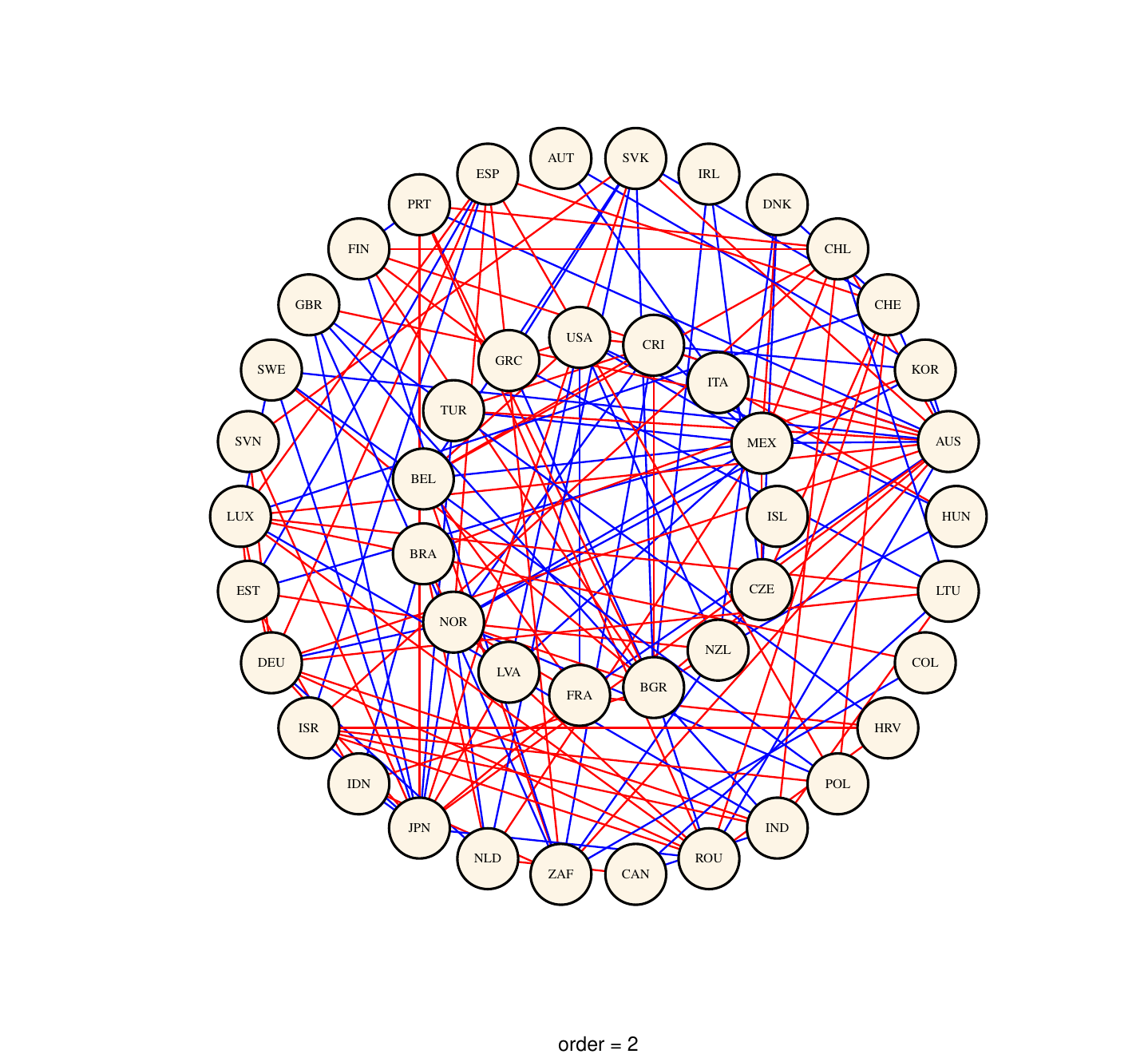}
        \caption{Shifts in total GDP, highlighting overall economic impact; using spOUTAR with order 2.}
        \label{fig:gdp}
    \end{subfigure}

    \vspace{0.5cm} 

    \begin{subfigure}[b]{0.40\textwidth}
        \centering
        \includegraphics[width=\textwidth,trim=4.4cm 3.2cm 3.3cm 2.5cm,clip]{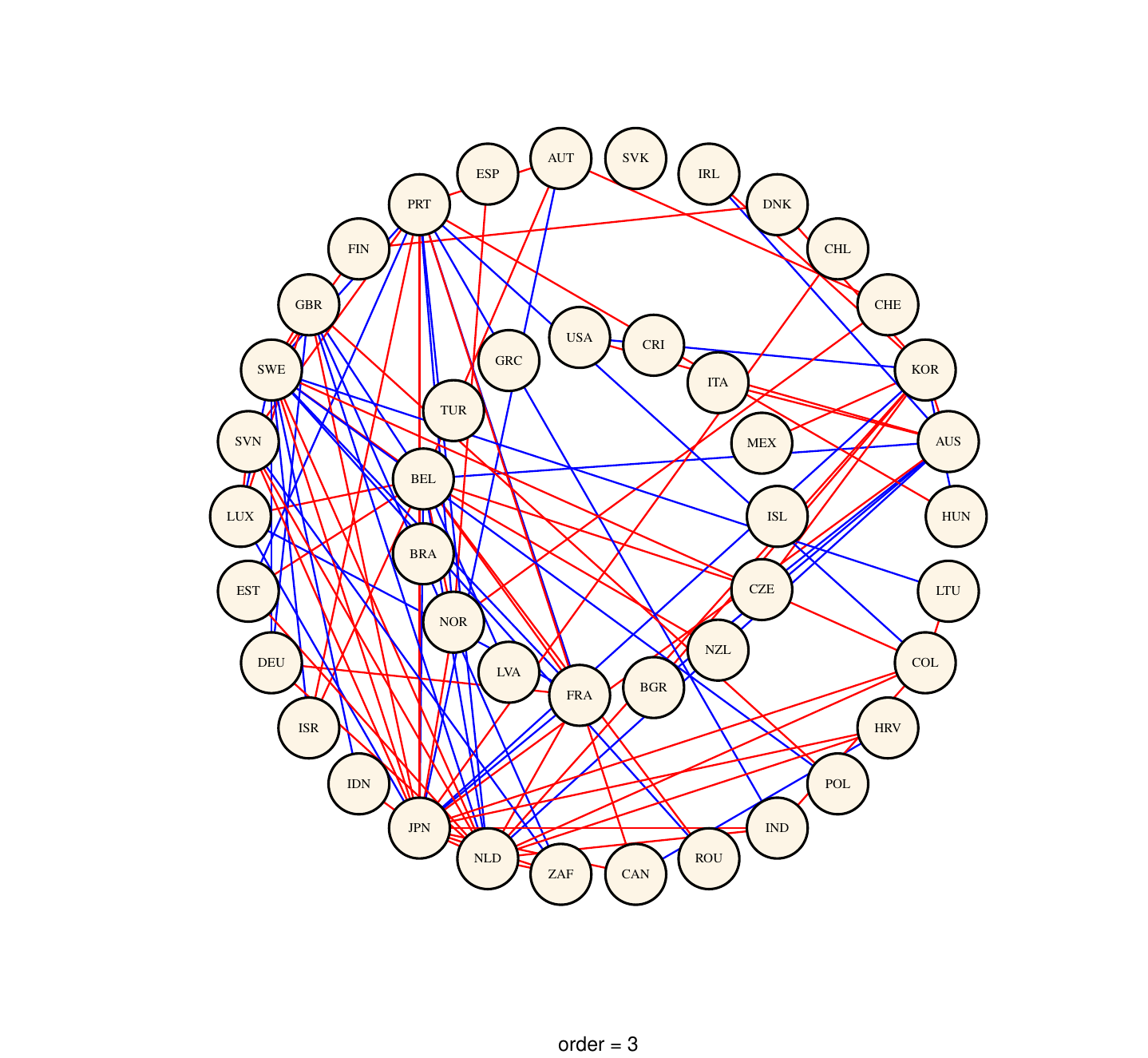}
        \caption{Changes in household and non-profit organization expenditures, showing consumption adjustments; using spOUTAR with order 3.}
        \label{fig:housexp}
    \end{subfigure}
    \hfill
    \begin{subfigure}[b]{0.40\textwidth}
        \centering
        \includegraphics[width=\textwidth,trim=4.4cm 3.2cm 3.3cm 2.5cm,clip]{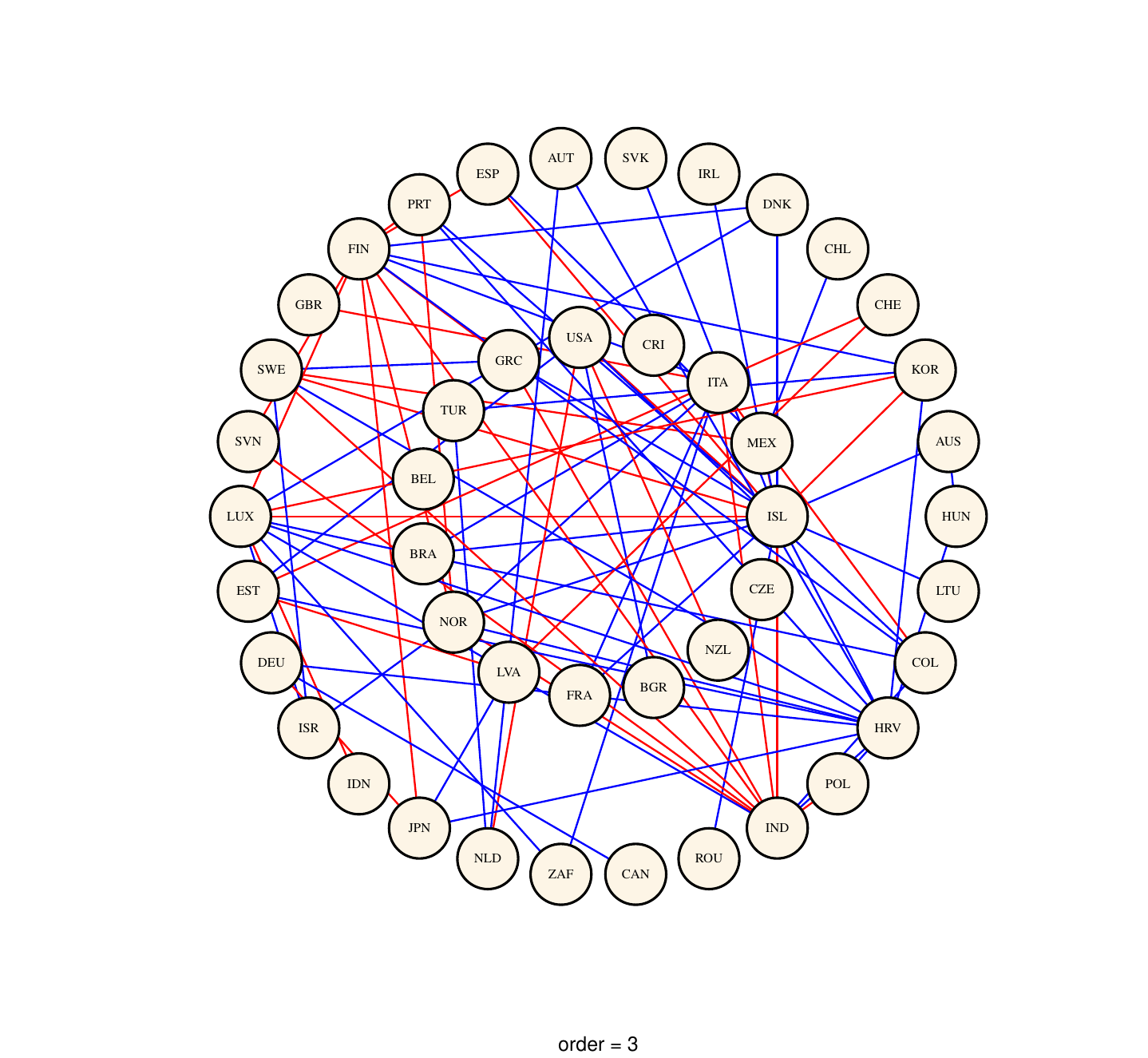}
        \caption{Changes in capital investment and fixed asset indicators, illustrating investment trends; using spOUTAR with order 3.}
        \label{fig:capital}
    \end{subfigure}

    \caption{Post-2009 recession structural shifts in: (a) government expenditure, (b) household/non-profit expenditure, (c) total GDP, and (d) capital investments.}
    \label{fig:oecd_2x2}
\end{figure}

We report some additional results related to Section \ref{sec:real} in this section. The general government expenditure of a country primarily comprises individual expenditures: spending on services provided to specific individuals, such as healthcare or education, and collective expenditures: spending on services benefiting the community as a whole, like defense, justice, or public administration. In Figure~\ref{fig:govexp}, we can observe a rich dependency structure in the government expenditure of OECD countries, consisting of many positive and negative shifts. Such changes in relationships indicate the formation of new trade agreements, changes to existing trade policies, and similar welfare systems, giving rise to new trade partnerships and decreasing reliance on traditional markets.

\newpage

\begin{figure}
    \centering
    \includegraphics[width=\textwidth,trim=2.0cm 0.8cm 2.0cm 1.6cm,clip]{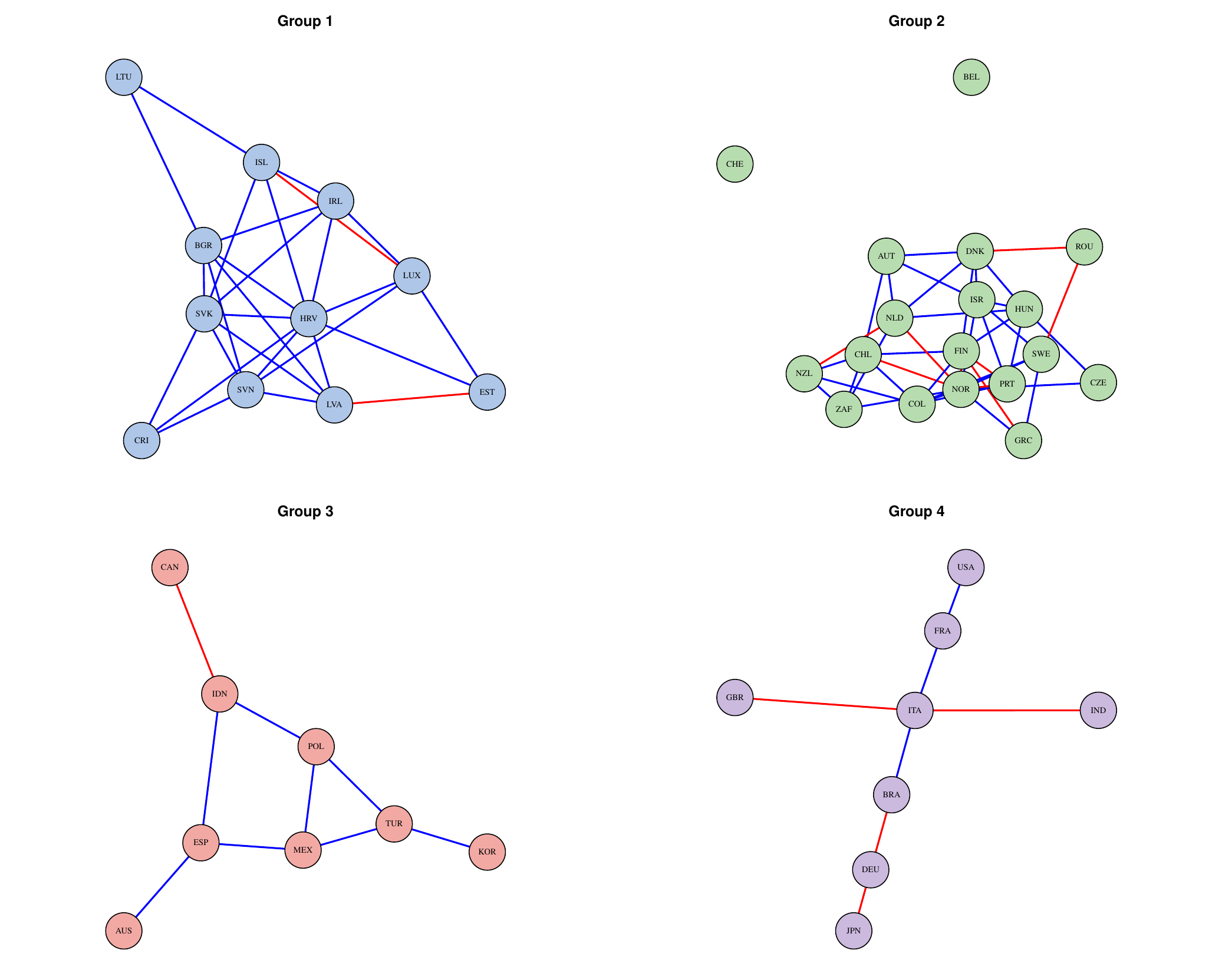}
    \caption{Group-wise classification of shifts in capital expenditures among the OECD nations following the recession as obtained from spOUTAR with order 3.}
    \label{fig:placeholder}
\end{figure}

\begin{figure}
    \centering
    \includegraphics[width=\textwidth,trim=2.0cm 0.8cm 2.0cm 1.6cm,clip]{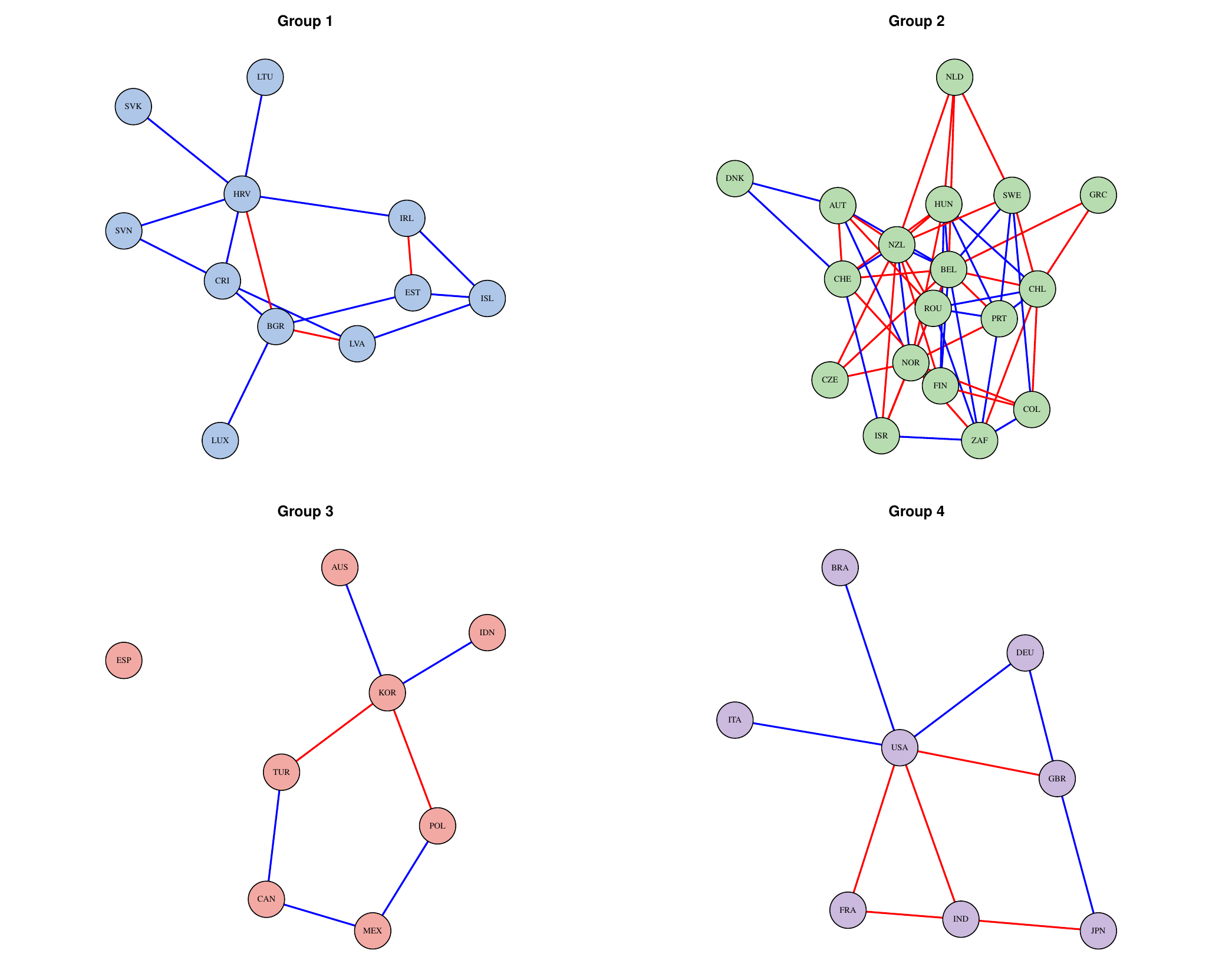}
    \caption{Group-wise classification of shifts in government expenditures among the OECD nations following the recession as obtained from spOUTAR with order 4.}
    \label{fig:placeholder}
\end{figure}

\begin{figure}
    \centering
    \includegraphics[width=\textwidth,trim=2.0cm 0.8cm 2.0cm 1.6cm,clip]{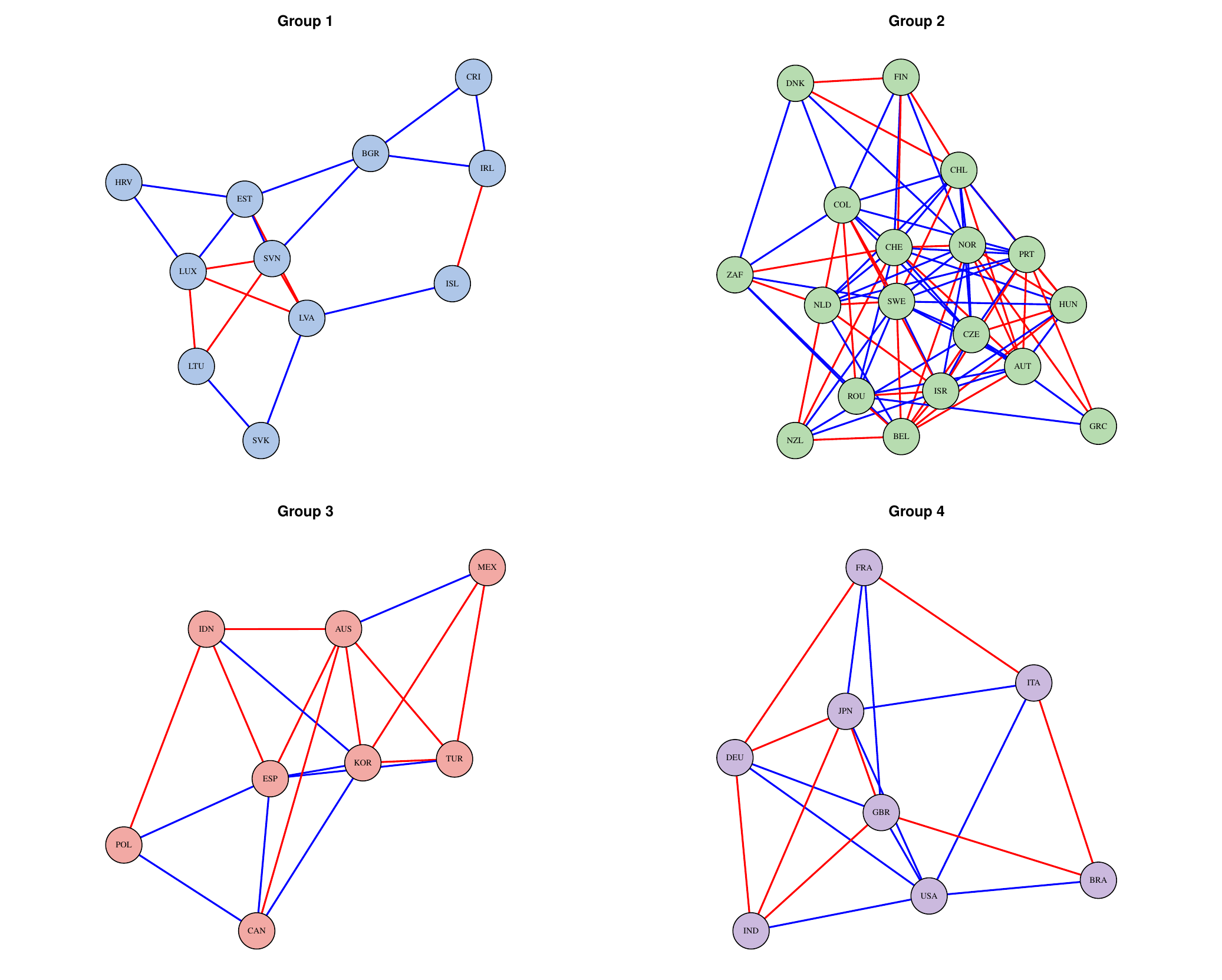}
    \caption{Group-wise classification of shifts in household and non-profit expenditures among the OECD nations following the recession as obtained from spOUTAR with order 3.}
    \label{fig:placeholder}
\end{figure}

\subsection{Supplementary results: simulated data analysis}\label{sec:appsim}
The comparison of performance of GGM, GCGM, and spOUTAR, for the $75\%$ sparsity level of $\Om_0$ is displayed in Table~\ref{tab:sim1}, whereas that for sparsity level $95\%$ is shown in Table~\ref{tab:sim3}.
\begin{table}[h!]
\centering
\resizebox{0.70\textwidth}{!}{
\begin{tabular}{@{}llcrrrllcrrr@{}}
\toprule
\text{$p$} & \text{$n$} & \text{$q$} & \text{GGM} & \text{GCGM} & \text{spOUTAR} &\text{$p$} & \text{$n$} & \text{$q$} & \text{GGM} & \text{GCGM} & \text{spOUTAR} \\ \midrule
30 & 50  & 2 & 5.7663 & \textbf{2.7016} & 3.5120 & 30 & 150 & 8 & 26.1881 & 5.5204 & \textbf{3.9208} \\
30 & 100 & 2 & 9.8990 & \textbf{2.8385} & 3.9385 & 30 & 200 & 8 & 33.7859 & 7.3007 & \textbf{3.8094} \\
30 & 150 & 2 & 13.0946 & \textbf{3.0159} & 3.9237 & 30 & 250 & 8 & 41.0031 & 7.2416 & \textbf{3.9751} \\
60 & 50  & 2 & 4.8553 & \textbf{3.0263} & 4.0939 & 60 & 150 & 8 & 18.7041 & 4.9537 & \textbf{4.5559} \\
60 & 100 & 2 & 7.8095 & \textbf{3.0808} & 4.3947 & 60 & 200 & 8 & 24.6697 & 6.2315 & \textbf{4.4261} \\
60 & 150 & 2 & 10.5585 & \textbf{3.2697} & 4.4669 & 60 & 250 & 8 & 29.9310 & 6.0104 & \textbf{4.4599} \\
90 & 50  & 2 & 4.9125 & \textbf{2.7682} & 3.4992 & 90 & 150 & 8 & 16.8037 & 4.7577 & \textbf{3.6718} \\
90 & 100 & 2 & 7.5720 & \textbf{2.7268} & 3.6915 & 90 & 200 & 8 & 22.1458 & 5.8282 & \textbf{3.5927} \\
90 & 150 & 2 & 9.7087 & \textbf{2.5876} & 3.6159 & 90 & 250 & 8 & 25.8131 & 6.2915 & \textbf{3.9384} \\
\midrule
30 & 50  & 5 & 8.3758 & \textbf{2.9174} & 3.7243 & 30 & 150 & 10 & 27.1034 & 6.7129 & \textbf{3.8547} \\
30 & 100 & 5 & 15.5352 & \textbf{3.4515} & 3.8630 & 30 & 200 & 10 & 35.3740 & 8.7824 & \textbf{3.8439} \\
30 & 150 & 5 & 22.0368 & 4.0451 & \textbf{3.9490} & 30 & 250 & 10 & 44.1561 & 9.4918 & \textbf{3.8736} \\
60 & 50  & 5 & 6.4550 & \textbf{3.2296} & 4.2286 & 60 & 150 & 10 & 19.8073 & 5.6529 & \textbf{4.4585} \\
60 & 100 & 5 & 11.4006 & \textbf{3.5430} & 4.4444 & 60 & 200 & 10 & 25.9693 & 7.2165 & \textbf{4.3956} \\
60 & 150 & 5 & 16.1652 & \textbf{3.7312} & 4.4626 & 60 & 250 & 10 & 31.8303 & 7.4142 & \textbf{4.4025} \\
90 & 50  & 5 & 6.3952 & \textbf{3.1627} & 3.5848 & 90 & 150 & 10 & 8.9084 & 3.5498 & \textbf{1.8847} \\
90 & 100 & 5 & 10.5872 & \textbf{3.4964} & 3.7118 & 90 & 200 & 10 & 22.7985 & 6.5974 & \textbf{3.7353} \\
90 & 150 & 5 & 14.7463 & 3.7244 & \textbf{3.5480} & 90 & 250 & 10 & 27.7146 & 8.1353 & \textbf{3.7046} \\
\bottomrule
\end{tabular}
}
\caption{Correctly-specified AR case: Comparison of average RMSE value in the precision matrix estimation where the true precision is $75\%$ sparse, following a small world networks from the Watts-Strogatz model for different choices of dimension $p$, sample size $n$, and AR order $q$.}
\label{tab:sim1}
\end{table}

\begin{table}[h!]
\centering
\resizebox{0.70\textwidth}{!}{
\begin{tabular}{@{}llcrrrllcrrr@{}}
\toprule
\text{$p$} & \text{$n$} & \text{$q$} & \text{GGM} & \text{GCGM} & \text{spOUTAR} &\text{$p$} & \text{$n$} & \text{$q$} & \text{GGM} & \text{GCGM} & \text{spOUTAR} \\ \midrule
30 & 50  & 2 & 6.1040 & \textbf{2.1191} & 2.5444 & 30 & 150 & 8 & 26.7892 & 5.7238 & \textbf{2.8982} \\
30 & 100 & 2 & 10.3034 & \textbf{2.2287} & 2.8205 & 30 & 200 & 8 & 34.4414 & 6.6932 & \textbf{2.9884} \\
30 & 150 & 2 & 13.2441 & \textbf{2.3665} & 2.7987 & 30 & 250 & 8 & 41.9731 & 6.8604 & \textbf{2.8952} \\
60 & 50  & 2 & 5.8100 & 1.7967 & \textbf{1.7293} & 60 & 150 & 8 & 20.4444 & 5.1302 & \textbf{2.1133} \\
60 & 100 & 2 & 8.7152 & \textbf{1.7201} & 2.1147 & 60 & 200 & 8 & 25.9356 & 5.4493 & \textbf{2.1940} \\
60 & 150 & 2 & 11.3598 & \textbf{1.7186} & 2.1767 & 60 & 250 & 8 & 31.4138 & 6.1560 & \textbf{2.0539} \\
90 & 50  & 2 & 5.2888 & 1.8296 & \textbf{1.6129} & 90 & 150 & 8 & 17.5272 & 4.8658 & \textbf{1.5902} \\
90 & 100 & 2 & 7.9315 & \textbf{1.5605} & 1.5725 & 90 & 200 & 8 & 22.6506 & 5.4356 & \textbf{1.7678} \\
90 & 150 & 2 & 10.4298 & \textbf{1.4095} & 1.9111 & 90 & 250 & 8 & 26.7136 & 6.6242 & \textbf{1.8346} \\
\midrule
30 & 50  & 5 & 9.0106 & \textbf{2.5357} & 2.6028 & 30 & 150 & 10 & 27.7238 & 7.7051 & \textbf{2.7569} \\
30 & 100 & 5 & 16.1381 & 3.0613 & \textbf{2.8607} & 30 & 200 & 10 & 36.3130 & 7.4012 & \textbf{2.8068} \\
30 & 150 & 5 & 22.8368 & 3.7322 & \textbf{2.8863} & 30 & 250 & 10 & 44.3195 & 8.9971 & \textbf{2.8235} \\
60 & 50  & 5 & 7.6580 & 2.7688 & \textbf{1.9800} & 60 & 150 & 10 & 21.1959 & 6.2402 & \textbf{2.1510} \\
60 & 100 & 5 & 12.8130 & 2.9516 & \textbf{2.1114} & 60 & 200 & 10 & 27.0981 & 7.9321 & \textbf{2.3463} \\
60 & 150 & 5 & 17.5129 & 3.1878 & \textbf{2.0956} & 60 & 250 & 10 & 33.1425 & 9.0242 & \textbf{2.1189} \\
90 & 50  & 5 & 6.9174 & 3.0060 & \textbf{1.4958} & 90 & 150 & 10 & 32.1654 & 7.0407 & \textbf{1.4259} \\
90 & 100 & 5 & 11.3200 & 2.6949 & \textbf{1.8013} & 90 & 200 & 10 & 30.9774 & 8.1224 & \textbf{1.3153} \\
90 & 150 & 5 & 14.9809 & 2.4116 & \textbf{1.7804} & 90 & 250 & 10 & 28.2674 & 6.7610 & \textbf{1.9941} \\
\bottomrule
\end{tabular}
}
\caption{Correctly-specified AR case: Comparison of average RMSE value in the precision matrix estimation where the true precision is $95\%$ sparse, following a small world networks from the Watts-Strogatz model for different choices of dimension $p$, sample size $n$, and AR order $q$.}
\label{tab:sim3}
\end{table}

\end{document}